\documentclass[fleqn,usenatbib]{mnras}

\usepackage{newtxtext,newtxmath}

\usepackage[T1]{fontenc}

\DeclareRobustCommand{\VAN}[3]{#2}
\let\VANthebibliography\thebibliography
\def\thebibliography{\DeclareRobustCommand{\VAN}[3]{##3}\VANthebibliography}

\usepackage{graphicx}	
\usepackage{amsmath}	

\title[Molecular hydrogen excess]{The excess of molecular hydrogen in chemical networks without oxygen}

\author[S. Ploeckinger]{
Sylvia Ploeckinger$^{1}$\thanks{E-mail: sylvia.ploeckinger@univie.ac.at}
\\
$^{1}$Department of Astrophysics, University of Vienna, T\"urkenschanzstrasse 17, 1180 Vienna, Austria
}

\date{Accepted XXX. Received YYY; in original form ZZZ}

\pubyear{\the\year{}}

\begin{document}
\label{firstpage}
\pagerange{\pageref{firstpage}--\pageref{lastpage}}
\maketitle

\begin{abstract}
We report the presence of a systematic excess in the molecular hydrogen fraction ($f_{\mathrm{H2}} = 2 \, n_{\mathrm{H2}}/n_{\mathrm{H}}$) in studies that use a reduced chemistry network to calculate $f_{\mathrm{H2}}$ of gas with a non-zero metal mass fraction. This is common practice in simulations of galaxy formation in which following the non-equilibrium abundances of additional elements is computationally expensive. We define the $\mathrm{H}_2$ excess as the shift in density of the \ion{H}{I}-$\mathrm{H}_2$ transition in the reduced network compared to the full chemical network (30 elements). The strength of the $\mathrm{H}_2$ excess generally increases both with temperature and metallicity, is largely independent of the radiation field strength, and persists across a large range of assumed shielding column densities. For warm gas, with $T\approx1000~\mathrm{K}$, the \ion{H}{I}-$\mathrm{H}_2$ transition is shifted by up to 1~dex to lower densities in primordial chemistry networks already for extremely low metallicities ($Z\geq 10^{-4}\,\mathrm{Z}_{\odot}$).  
We confirm our earlier findings that missing reactions with oxygen are largely responsible for this $\mathrm{H}_2$ excess. A reduced chemical network of hydrogen, helium, and oxygen recovers the molecular hydrogen fractions from a full network and we therefore recommend to include destruction of molecular hydrogen by oxygen in a minimal chemical network for accurate molecular hydrogen abundances. 


\end{abstract}




\section{Introduction}\label{sec:introduction}

Molecular hydrogen, $\mathrm{H}_2$, is the most abundant molecule in the Universe and a critical ingredient for many astrophysical processes. In gas with extremely low metallicities, as typical for the formation of stars in the very early Universe, $\mathrm{H}_2$ is the main radiative cooling process for temperatures, $T$, below $10^4\,\mathrm{K}$, as metal-line cooling is inefficient or absent. In the interstellar medium (ISM) of metal-rich galaxies, such as the Milky Way, $\mathrm{H}_2$ is the primary component of dense molecular clouds. Together with metal-line cooling, it contributes to the total radiative cooling of neutral gas, allowing the gas to reach lower temperatures and increasing the likelihood of fragmentation into smaller regions, which ultimately leads to the formation of individual stars.

Direct observation of $\mathrm{H}_2$ emission is challenging (see \citealp{Wakelam2017} for a review). Because of its symmetric structure, $\mathrm{H}_2$ does not emit or absorb radiation through dipole transitions, which are the most common and easily detectable transitions in molecules. In warm gas ($T \approx 1000\,\mathrm{K}$), collisions can excite $\mathrm{H}_2$ molecules to higher vibrational and rotational states. These excited states decay via quadrupole transitions, emitting infrared radiation and enabling radiative cooling. Only if a molecular cloud is heated by a strong radiation field, $\mathrm{H}_2$ can be directly observed via fluorescent emission, while it is also in the process of being photo-dissociated \citep{Burkhart2025}. Due to the short cooling time at these densities, the majority of $\mathrm{H}_2$ is in cold ($\lesssim 100\,\mathrm{K}$) gas. At these temperatures, $\mathrm{H}_2$ remains in its ground state and does not emit radiation because quadrupole transitions require higher excitation energies (for $\mathrm{H}_2$: $T\ge 511\,\mathrm{K}$).

The contrast between the theoretical simplicity of molecular hydrogen and the observational challenges it presents has led to distinct approaches in studying molecular gas in the Universe. Observational studies rely on indirect tracers, such as CO, to infer the presence and distribution of $\mathrm{H}_2$. On the other hand, theoretical work often uses chemical networks that only contain a small subset of species (e.g. only hydrogen and helium species) and directly model the formation and abundance of $\mathrm{H}_2$ without providing information on the CO content for computational efficiency. A direct validation of theoretical studies on molecular gas through comparisons with a wide variety of observations is therefore difficult.

In \citet{Ploeckinger2025} we introduced \textsc{Hybrid-Chimes}, which is based on the chemical network solver \textsc{Chimes} from \citet{chimes2014optthin, chimes2014shielded} and implemented in the gravity and hydrodynamics code \textsc{Swift} \citep{swift2024}. For computational efficiency, \textsc{Hybrid-Chimes} allows the user to include any selection from metal elements C, N, O, Ne, Si, Mg, S, Ca, and Fe in addition to the primordial elements H and He. While the abundance ratios of the selected elements are evaluated in non-equilibrium and ``on-the-fly'', the abundance ratios as well as the cooling and heating rates from the remaining elements are pre-tabulated under the assumption of steady-state chemistry and ionization equilibrium\footnote{We use ``chemical equilibrium'' as short form to steady-state chemistry and ionization equilibrium throughout this work.}. For example, in the flagship simulations from the \textsc{Colibre} project \citep{Schaye2025colibre, Chaikin2025a}, ten hydrogen and helium species ($e^-$, $\mathrm{H}$, $\mathrm{H}^+$, $\mathrm{H}^-$, $\mathrm{He}$, $\mathrm{He}^+$, $\mathrm{He}^{++}$, $\mathrm{H}_2$, $\mathrm{H}_2^+$, and $\mathrm{H}_3^+$) are evolved in non-equilibrium and their cooling and heating rates are added to those pretabulated from the metal elements (corrected for the non-equilibrium electron abundances, see \citealp{Ploeckinger2025} for details). 

For the specific assumptions on the radiation field, cosmic ray rate, shielding column density and dust content in \citet{Ploeckinger2025}, we found a systematic over-estimation of $\mathrm{H}_2$ when using the minimal chemical network (only H and He species), compared to the results from the full chemical network (11 elements, 157 species). We showed that by adding oxygen to the chemical network (H, He, and O species), the $\mathrm{H}_2$ fraction from the full chemical network is recovered. In the following, we discuss some examples to demonstrate that reactions with oxygen species are commonly excluded for the calculation of the abundance of $\mathrm{H}_2$. 

\textsc{Grackle} \citep{Grackle2017} is a widely used chemical network, implemented e.g. in the gravity / hydro-dynamics codes ChaNGa \citep{Changa2015}, Cholla \citep{Cholla2015}, Enzo \citep{Enzo2014}, Gasoline \citep{Gasoline2017}, Gizmo \citep{Gizmo2015}, Ramses \citep{Teyssier2002}, Arepo \citep{Arepo2020} and SWIFT \citep{swift2024}, 
which solves the non-equilibrium evolution of primordial elements (hydrogen and helium species). 
In order to account for cooling and heating processes from metal species, especially metal-line cooling, the contributions of metals are calculated and pre-tabulated with \textsc{Cloudy} (last described in \citealp{cloudy2025}), assuming chemical equilibrium. Examples for this approach include the AGORA galaxy simulation comparison project \citep{Agora2014}, simulations of isolated galaxies \citep{Emerick2018, Emerick2019, Smith2021}, cosmological zoom-in simulations of dwarf galaxies \citep{Revaz2018}, and patches of the interstellar medium \citep{Gentry2019}. In the cosmological periodic volume simulation \textsc{Simba-EOR} \citep{SimbaEOR2024}, the molecular hydrogen fraction is also calculated by \textsc{Grackle}'s primordial chemistry network\footnote{In earlier projects (\textsc{Mufasa}: \citealp{Dave2016} and \textsc{Simba}: \citealp{Dave2019}), the $\mathrm{H}_2$ fractions are taken from a subgrid model based on \citet{Krumholz2009}.}.
In all these examples, the formation and destruction of molecular hydrogen does not include reactions with gas-phase metals. 

The \textsc{Silcc} project \citep{Walch2015} uses a chemical network that follows the evolution of seven chemical species: $e^-$, $\mathrm{H}$, $\mathrm{H}^+$, $\mathrm{H}_2$, $\mathrm{C}^+$, $\mathrm{O}$, and $\mathrm{CO}$ and is implemented in the adaptive mesh refinement code \textsc{Flash} \citep{Fryxell2000} to model the evolution of the interstellar medium. While the evolution of some metal species is included in the chemical network, the reactions that form and destroy molecular hydrogen (full list in \citealp{Micic2012}) do not include reactions with metal species. The reactions listed in \citet{Micic2012} are also used in \citet{Hu2016,Hu2017Grackle,Hu2021} and the \textsc{Griffin} simulations \citep{Lahen2020}, again missing the reactions between $\mathrm{H}_2$ and metal species. The chemical network \textsc{Hyacinth} \citep{Hyacinth2024} follows selected H, He, C, and O species. Here, $\mathrm{H}_2$ can be destroyed by reactions with $\mathrm{C}^+$, but not through reactions with oxygen species. 

The simulation project \textsc{Ramses-RTZ} \citep{Katz2022} follows the non-equilibrium species abundances of 8 metal elements (C, N, O, Mg, Si, Si, Fe, and Ne) in addition to H and He, coupled to the radiation hydrodynamics solver present in \textsc{Ramses-RT} \citep{Teyssier2002, Rosdahl2015}. While they follow the evolution of $\mathrm{H}_2$, only reactions of hydrogen and helium species with $\mathrm{H}_2$ are included, as described in \citet{Katz2017}. They benchmark \textsc{Ramses-RTZ} against the results from \textsc{Cloudy} in \citet{Katz2022} but exclude molecules for this comparison. 

\textsc{Thesan-Zoom} \citep{Kannan2025} follows individual haloes until $z=3$ with the non-equilibrium chemical network from \citet{Kannan2020} which evolves the abundances of hydrogen and helium species. They include radiative transfer, particularly of Lyman-Werner radiation, and study e.g. the formation of Pop III stars \citep{Zier2025} and the transition to the formation of Pop II stars. For these low metallicities, metal-line cooling is reduced and cooling from $\mathrm{H}_2$ is the dominant cooling process for gas with temperatures of $T\lesssim 10^4\,\mathrm{K}$. The destruction of $\mathrm{H}_2$ is modeled with collisions of $\mathrm{H}_2$ molecular with atomic hydrogen and other hydrogen molecules, as well as photo-ionization and photo-dissociation of $\mathrm{H}_2$, but does not include reactions with metal species. 

In this paper, we investigate if the $\mathrm{H}_2$ excess found in \citet{Ploeckinger2025} is limited to the specific model assumptions (radiation field, shielding column density, cosmic ray rate, gas metallicity) and chemical network (\textsc{Chimes}) or if this may indicate a general, systematic overestimation of $\mathrm{H}_2$ in reduced chemical networks that include a subset of species and in particular do not include reactions of $\mathrm{H}_2$ and oxygen species. We present the method in section~\ref{sec:method} and demonstrate the ubiquity of the $\mathrm{H}_2$ excess in section~\ref{sec:results}. In this section (section~\ref{sec:reducingH2excess}) we also show that the $\mathrm{H}_2$ excess is drastically reduced when including reactions with oxygen. We discuss implications for studies of both low and high metallicity gas in sections~\ref{sec:lowZ} and \ref{sec:highZ} and provide a brief summary in section~\ref{sec:summary}. Throughout the paper $\log$ refers to $\log_{\mathrm{10}}$.

\section{Method}\label{sec:method}

We use \textsc{Cloudy} version 25.00, last described in \citet{cloudy2025}, to calculate the fraction of hydrogen in molecular hydrogen, $f_{\mathrm{H2}}\equiv 2\,n_{\mathrm{H2}} / n_{\mathrm{H}}$, with the molecular hydrogen number density, $n_{\mathrm{H2}}$, and the total hydrogen number density, $n_{\mathrm{H}}$. 

For each \textsc{Cloudy} calculation, we assume that a slab of gas with constant density, temperature, and composition (i.e. element abundances and dust grain properties) is illuminated on one side in a plane-parallel geometry. \textsc{Cloudy} divides the gas into zones with an adaptive thickness. Each subsequent zone is more shielded from the radiation field, as the depth into the cloud, and therefore the shielding column density, $N_{\mathrm{sh}}$, increases. We stop the calculations at a maximum total hydrogen shielding column density, $N_{\mathrm{sh,max}}$, and record the molecular hydrogen fraction, $f_{\mathrm{H2}}$ for each zone. By default, \textsc{Cloudy} evaluates the full gas column once, but the correct total optical depths may be important for line transfer, so we use the option to iterate the calculations of the full shielding column until the optical depths of the total gas column have converged\footnote{In rare occasions, a phase transition (e.g. an ionization or dissociation front) coincides with the end of the gas column, which can lead to large relative changes of the total optical depth for each iteration and \textsc{Cloudy} does not find a converged result. See \textsc{Cloudy} documentation for details.}. The  $\mathrm{H}_2$ chemistry in \textsc{Cloudy} is described in detail in \citet{Shaw2005}. 

We use $\log N_{\mathrm{sh,max}} \, [\mathrm{cm}^{-2}] = 22.5$, which ensures that the typical shielding column densities for the \ion{H}{I}-$\mathrm{H}_2$ transition in the local Universe (e.g. Milky Way Galaxy: $\log N_{\mathrm{H,HI\rightarrow H_2}} \, [\mathrm{cm}^{-2}] \approx 21.4$, \citealp{Wolfire2008}; Small Magellanic Cloud: $\log N_{\mathrm{H,HI\rightarrow H_2}} \, [\mathrm{cm}^{-2}] \approx 22.3$, \citealp{Bolatto2011}) are included for reference. A larger maximum shielding column density would not affect our analysis on the \ion{H}{I}-$\mathrm{H}_2$ transition, because hydrogen is typically fully molecular at these column densities. 

\begin{table}
    \caption{Chemical networks used in this work. Independent of the included elements, the same dust composition is included in each chemical network. This reproduces the approach of minimal chemical networks to calculate $\mathrm{H}_2$ fractions.}
    \label{tab:chemicalnetworks}
    \begin{tabular}{l|c|p{3.5cm}}
    \hline
    Network label & Number of elements & Elements\\
    \hline
    \texttt{HHe} & 2  & H, He\\
    \texttt{HHeO} & 3 & H, He, O \\
    \texttt{HHeC} & 3 & H, He, C \\
    \texttt{HHeCO} & 4 & H, He, C, O \\
    \texttt{Full}&  30 & H, He, Li, Be, B, C, N, O, F, Ne, Na, Mg, Al, Si, P, S, Cl, Ar, K, Ca, Sc, Ti, V, Cr, Mn, Fe, Co, Ni, Cu, Zn\\
    \hline
    \end{tabular}
\end{table}

For the full chemical network, \texttt{Full}, we use the solar abundance ratios from \citet{Asplund2009} for all 30 elements included in \textsc{Cloudy} (see Table~\ref{tab:chemicalnetworks}) and scale the total abundance of elements heavier than helium with the relative metallicity $Z/Z_{\odot}$, with $Z_{\odot} = 0.0134$ \citep{Asplund2009}. Dust grains (silicates and graphites) and PAHs are included with a size distribution that reproduces the overall extinction properties observed in the ISM of the Milky Way Galaxy. This dust model has a dust-to-gas ratio of $\mathcal{DTG} = 6.6\times 10^{-3}$ for solar metallicity and is the default dust model in \textsc{Cloudy}. We assume a constant dust-to-metal ratio. Metals are depleted from the gas phase following the strong depletion model ($F_{\star} = 1$) from \citet{Jenkins2009}. While \textsc{Cloudy} does not check for consistency between the grain model and the gas-phase depletion model, we have shown in section 2.1.5 in \citet{Ploeckinger2025} that the $F_{\star} = 1$ depletion model from \citet{Jenkins2009} is consistent with the dust to gas ratio from \textsc{Cloudy}'s ISM dust grain model.  

The radiation field illuminating the gas and dust mixture consists of the cosmic microwave background at $z=0$ and an interstellar radiation field (ISRF) whose shape and intensity \citep{Black1987} is representative of the solar neighborhood, and combines the solar vicinity radiation field \citep{Mathis1983} with a Galactic soft X-ray background \citep{Bregman1986}. For reference, the strength of this ISRF is $n_{\gamma,\mathrm{ref}} = 3.7\times 10^{-4}\,\mathrm{photons\,cm}^{-3}$, where $n_{\gamma,\mathrm{ref}}$ is the number density of hydrogen ionizing photons ($> 13.6\,\mathrm{eV}$), or $u_{\mathrm{ref}} =  1.6\times10^{-3}\,\mathrm{erg\,cm}^{-2}\,\mathrm{s}^{-1}$, where  $u_{\mathrm{ref}}$ is the integrated energy flux in the energy range between 6 and 13.6~eV ($G_0 = 1$ in units of \citealp{Habing1968}).

Finally, a cosmic ray (CR) background is added with a mean hydrogen CR ionization rate of $\zeta_{\mathrm{ref}} =2\times 10^{-16}\,\mathrm{s}^{-1}$ \citep{Indriolo2015} and an $\mathrm{H}_2$ secondary ionization rate of $4.6\times10^{-16}\,\mathrm{s}^{-1}$, following \citet{Glassgold1973}.

\subsection{Model grid}\label{sec:modelgrid}

We run \textsc{Cloudy} grids in gas temperature, $T\,[\mathrm{K}]$, and gas density, $n_{\mathrm{H}}$, with different radiation field field strengths ($I_{\mathrm{ISRF}} = n_{\gamma} / n_{\gamma,\mathrm{ref}}  = u / u_{\mathrm{ref}}$), cosmic ray ionization rates ($I_{\mathrm{CR}} = \zeta_{\mathrm{CR}}/ \zeta_{\mathrm{ref}}$), and metallicities ($Z/Z_{\odot}$), each defined relative to their respective reference values. As both cosmic rays and the interstellar radiation field are related to star formation processes, we assume $I_{\mathrm{ISRF}} = I_{\mathrm{CR}}$. We evaluate each \textsc{Cloudy} calculation at various shielding column densities, $N_{\mathrm{sh}}$, for which we use the values of the zones closest to $N_{\mathrm{sh}}$. Table~\ref{tab:parameters} summarizes the parameter range and variations. For reproducibility, we provide the complete scripts to set up and analyze the \textsc{Cloudy} calculations, see the Data Availability Statement for more information.

 \begin{table}
  \caption{Model grid for the \textsc{Cloudy} calculations analyzed in this work. For each parameter we list the minimum (2nd column) and maximum (3rd column) values as well as the grid spacing (4th column).}
  \label{tab:parameters}
  \begin{tabular}{lccc}
    \hline
     & min & max & $\Delta$ \\
    \hline
    $\log T\,[\mathrm{K}]$   & 1.5 & 3.0 & 0.5 \\
    $\log n_{\mathrm{H}}\,[\mathrm{cm}^{-3}]$   & -3.0 & 4.5 & 0.05 \\
    $\log Z/Z_{\odot}$ &-4.0 & 0.0 & 1.0 \\
    $\log I_{\mathrm{ISRF}} = \log I_{\mathrm{CR}}$   & -2.0 & 1.0 & 1.0 \\
    $\log N_{\mathrm{sh}}\,[\mathrm{cm}^{-2}]$   & 19 & 22 & 0.5 \\
    \hline
  \end{tabular}
 \end{table}

\subsection{Chemical networks}

In order to quantify the excess of $\mathrm{H}_2$ in reduced networks, we run the model grid calculations for chemical networks with different sets of included elements. The full chemical network within \textsc{Cloudy} consists of 30 elements, their various ionization stages as well as their molecules. This is the reference model and we refer to it as \texttt{Full}. For the reduced chemical networks, we switch off all elements beside a small subset, always including H and He, by adding the \verb`element off [element name]` commands to the \textsc{Cloudy} input file. 

The dust grain composition in each reduced network matches the corresponding full chemical network comparison calculations. For solar metallicity, the formation on the surfaces of dust grains is the dominant formation process of $\mathrm{H}_2$, and therefore typically included in the reduced chemical networks mentioned in section~\ref{sec:introduction}. By excluding the individual elements from the gas-phase chemistry while keeping the grain properties unchanged, we reproduce the general processes included in reduced networks.

\begin{figure*}
 \includegraphics[width=\linewidth]{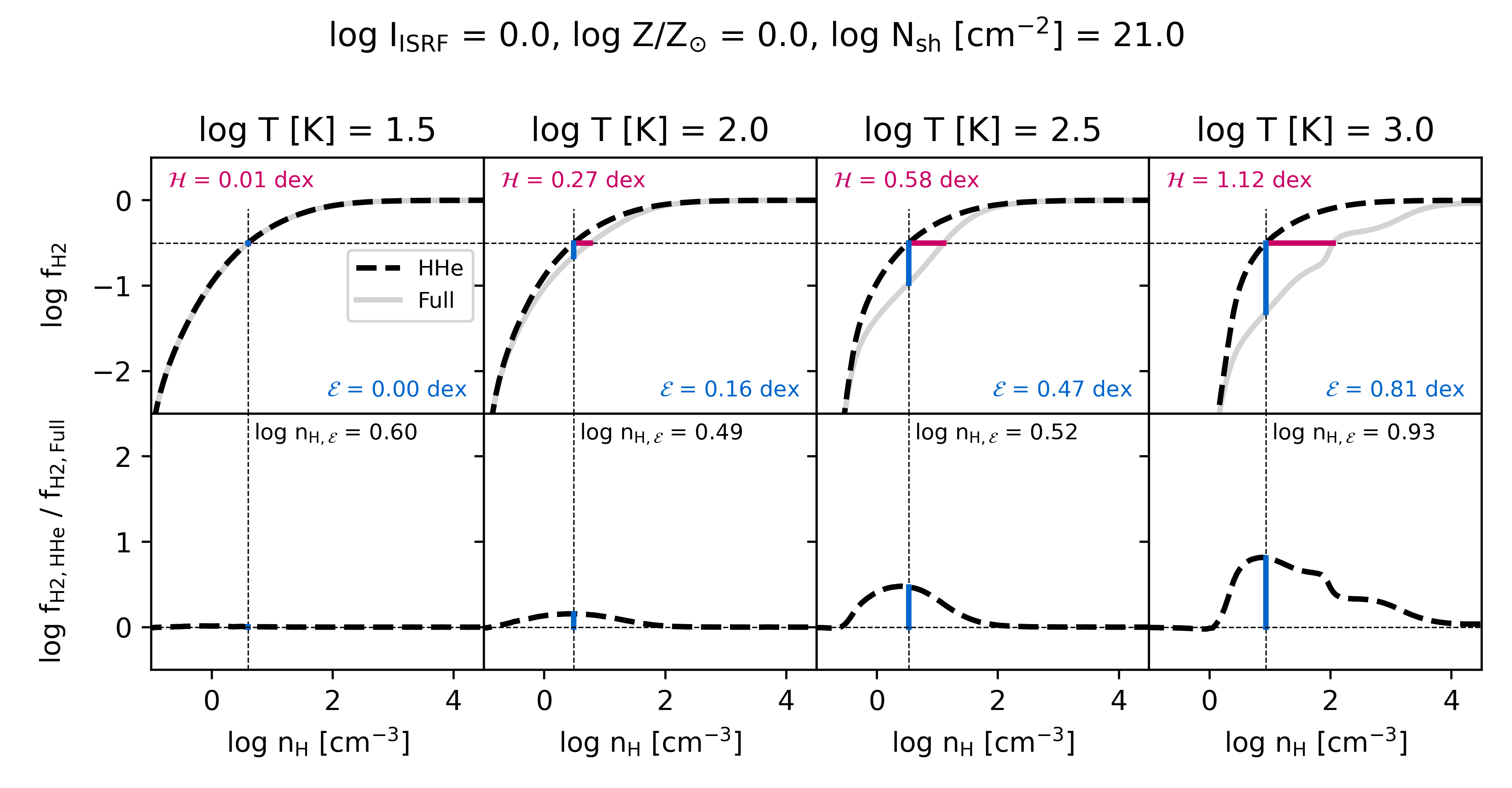}
 \caption{This figure illustrates the two definitions of the $\mathrm{H}_2$ excess, $\mathcal{H}$ (top row) and $\mathcal{E}$ (bottom row) for different gas temperatures (columns). The lines in the top row indicate the molecular hydrogen fractions, $\log f_{\mathrm{H2}} \equiv \log (2 n_{\mathrm{H2}} / n_{\mathrm{H}})$,  for total hydrogen densities, $n_{\mathrm{H}}$, for the reduced (H,He; dashed lines) and the full chemical network (H,He,M; solid lines). The thin horizontal line is at $\log f_{\mathrm{H2}} = - 0.5$, and serves as a marker for the transition between  atomic and molecular gas. The $\mathrm{H}_2$ excess is quantified as $\mathcal{H}$, the difference in the densities at which this transition occurs between the two networks (red horizontal bar) and $\mathcal{E}$, the difference in the $\mathrm{H}_2$ fraction at the density of the \ion{H}{I}-$\mathrm{H}_2$ transition in the reduced \texttt{HHe} network (indicated as blue vertical bar, also in the bottom panels). The black dashed line in the bottom panels shows $f_{\mathrm{H2}}$ from the reduced network \texttt{HHe} (black dashed line in top panel) relative to the $f_{\mathrm{H2}}$ from the full network (grey solid line in top panel). A thin horizontal line is added to the bottom panels at $f_{\mathrm{H2,HHe}} = f_{\mathrm{H2,Full}}$, i.e. no $\mathrm{H}_2$ excess, for reference. }
 \label{fig:H2excessindividual}    
\end{figure*}

\subsection{Definition of the $\mathrm{H}_2$ excess}

For each set of calculations, we quantify the over-estimation of $\mathrm{H}_2$ in the reduced networks, compared to the full network \texttt{Full}, by two different definitions of the $\mathrm{H}_2$ excess. The first one, $\mathcal{H}$, is defined as

\begin{equation}\label{eq:H}
    \mathcal{H} \, [\mathrm{dex}]= \log n_{\mathrm{H}}|_{\log f_{\mathrm{H2}}=-0.5} ^{\mathrm{Full}}\quad  -  \log n_{\mathrm{H}}|_{\log f_{\mathrm{H2}}=-0.5} ^{\mathrm{HHe}}
\end{equation}

\noindent
and describes the shift in density in dex at which the transition from atomic to molecular gas (i.e. the \ion{H}{I} to $\mathrm{H}_2$ transition) occurs, which we set at $\log f_{\mathrm{H2}} = -0.5$. The second quantity, $\mathcal{E}$, quantifies the ratio between the molecular hydrogen fractions from the reduced and the full network at the density at which $\log f_{\mathrm{H2}} = -0.5$ in the reduced network. 

The definitions and interpretations of $\mathcal{H}$ and $\mathcal{E}$ are illustrated in Fig.~\ref{fig:H2excessindividual}. The top row shows $f_{\mathrm{H2}}$ of the full (\texttt{Full}, grey solid lines) and the reduced (\texttt{HHe}, black dashed lines) chemical networks for different gas temperatures (columns). The thin dotted horizontal line indicates $\log f_{\mathrm{H2}} = -0.5$ which crosses the lines for both the full and the reduced networks. The distance in between is the $\mathrm{H}_2$ excess $\mathcal{H}$, highlighted as short horizontal red line, with the corresponding value in the top left of each panel. A value of $\mathcal{H} = 0.6 \,\mathrm{dex}$ means that the \ion{H}{I}-$\mathrm{H}_2$ transition is artificially shifted to lower densities, by 0.6~dex, when using a reduced network. The top row also demonstrates that while the value of $\mathcal{H}$ depends on the molecular hydrogen fraction that is chosen to represents the \ion{H}{I}-$\mathrm{H}_2$ transition (here: $\log f_{\mathrm{H2}} = -0.5$), the $\mathrm{H}_{2}$ excess extends over a wide range of $f_{\mathrm{H2}}$ and the general message of this work is unaffected by the exact choice of this value. 

\begin{figure}
    \centering
    \includegraphics[width=\linewidth]{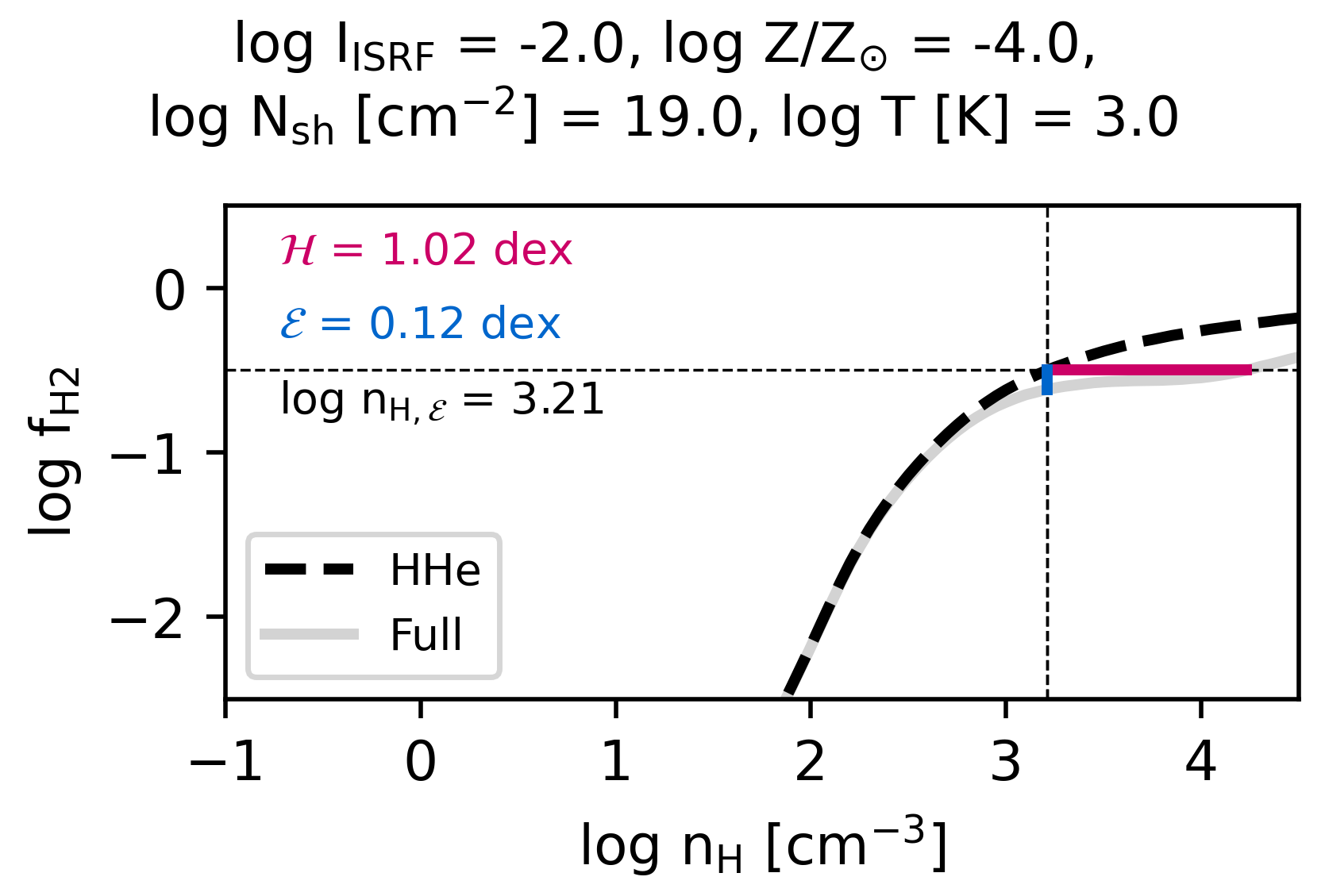}
    \caption{As top right panel in Fig.~\ref{fig:H2excessindividual} but for an example in which $\mathcal{E}\ll\mathcal{H}$ ($\log I_{\mathrm{ISRF}} = -2$, $\log Z/\mathrm{Z}_{\odot} = -4$, $\log N_{\mathrm{sh}} \, [\mathrm{cm^{-2}}] = 19$ and $\log T\,[\mathrm{K}] = 3$).}
    \label{fig:H2excess_single}
\end{figure}

The density at which $\log f_{\mathrm{H2}} = -0.5$ for the reduced chemical network \texttt{HHe} is defined as 

\begin{equation}\label{eq:n}
    \log n_{\mathrm{H},\mathcal{E}} \,[\mathrm{cm}^{-3}] = \log n_{\mathrm{H}}|_{\log f_{\mathrm{H2}}=-0.5} ^{\mathrm{HHe}} \quad,
\end{equation}

\noindent
indicated as thin vertical line, and the parameter $\mathcal{E}$, defined as 

\begin{eqnarray}\label{eq:E}
    \mathcal{E}  \, [\mathrm{dex}]=&& \log f_{\mathrm{H2}}^{\mathrm{HHe}}\, \left( \log n_{\mathrm{H},\mathcal{E}}\right) \nonumber\\
   &-& \log f_{\mathrm{H2}}^{\mathrm{CloudyFull}}\, \left(\log n_{\mathrm{H},\mathcal{E}} \right) \,,
\end{eqnarray}

\noindent
describes the error in the $\mathrm{H}_2$ fraction at density $\log n_{\mathrm{H},\mathcal{E}}$. $\mathcal{E}$ is represented as short vertical blue line in the top panel and the values for $\mathcal{E}$ and $\log n_{\mathrm{H},\mathcal{E}}$ are listed in the bottom right of each panel. For example, a value of $\mathcal{E} = 0.5\,\mathrm{dex}$ indicates that the $f_{\mathrm{H2}}$ fraction at the \ion{H}{I}-$\mathrm{H}_2$ transition in the reduced chemical network is 0.5~dex too high. The bottom row shows the ratio of the $\mathrm{H}_2$ fractions from the reduced networks to those from the full chemical network. The black dashed line is the ratio of the two lines ($f_{\mathrm{H2,HHe}}/ f_{\mathrm{H2,Full}}$) from the top panels and therefore the $\mathrm{H}_2$ excess parameter $\mathcal{E}$ can be directly seen here as well (vertical blue solid line). Depending on the steepness of the transition, $\mathcal{H}$ may be smaller (steep transition) or larger (shallow transition) than $\mathcal{E}$. In Fig.~\ref{fig:H2excess_single} we show an example for a slowly rising $\log f_{\mathrm{H2}}$ with $\log n_{\mathrm{H}}$, which results in a much smaller value for $\mathcal{E}$ than for $\mathcal{H}$ ($\mathcal{E} = 0.12\,\mathrm{dex}$, $\mathcal{H}=1.02\,\mathrm{dex}$).

\section{Result}\label{sec:results}

\begin{figure*}
    \centering
    \includegraphics[width=\linewidth]{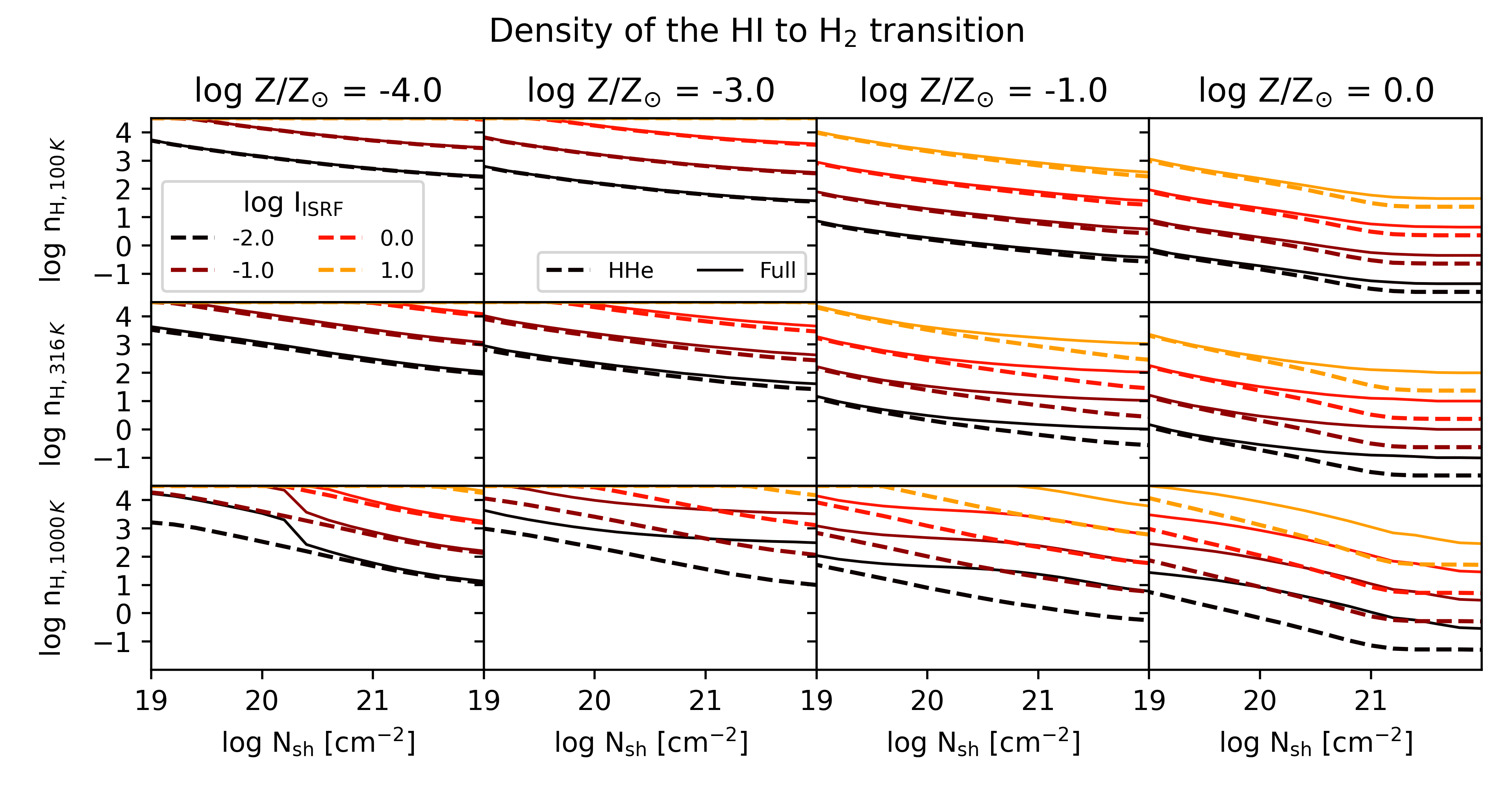}
    \caption{An overview of the densities of the \ion{H}{I}-$\mathrm{H}_2$ transition, defined as the density at which $\log f_{\mathrm{H2}} = -0.5$. Columns (metallicity), rows (temperature), and line colors (radiation field strength) as in Fig.~\ref{fig:H2excessH}. The solid lines show transition densities of the full chemical network, \texttt{Full}, and the dashed lines from the \texttt{HHe} network. The dashed lines therefore represent $\log n_{\mathrm{H},\mathcal{E}}$ from equation~\ref{eq:n}.}
    \label{fig:H2excessn}
\end{figure*}

\begin{figure*}
    \centering
    \includegraphics[width=\linewidth]{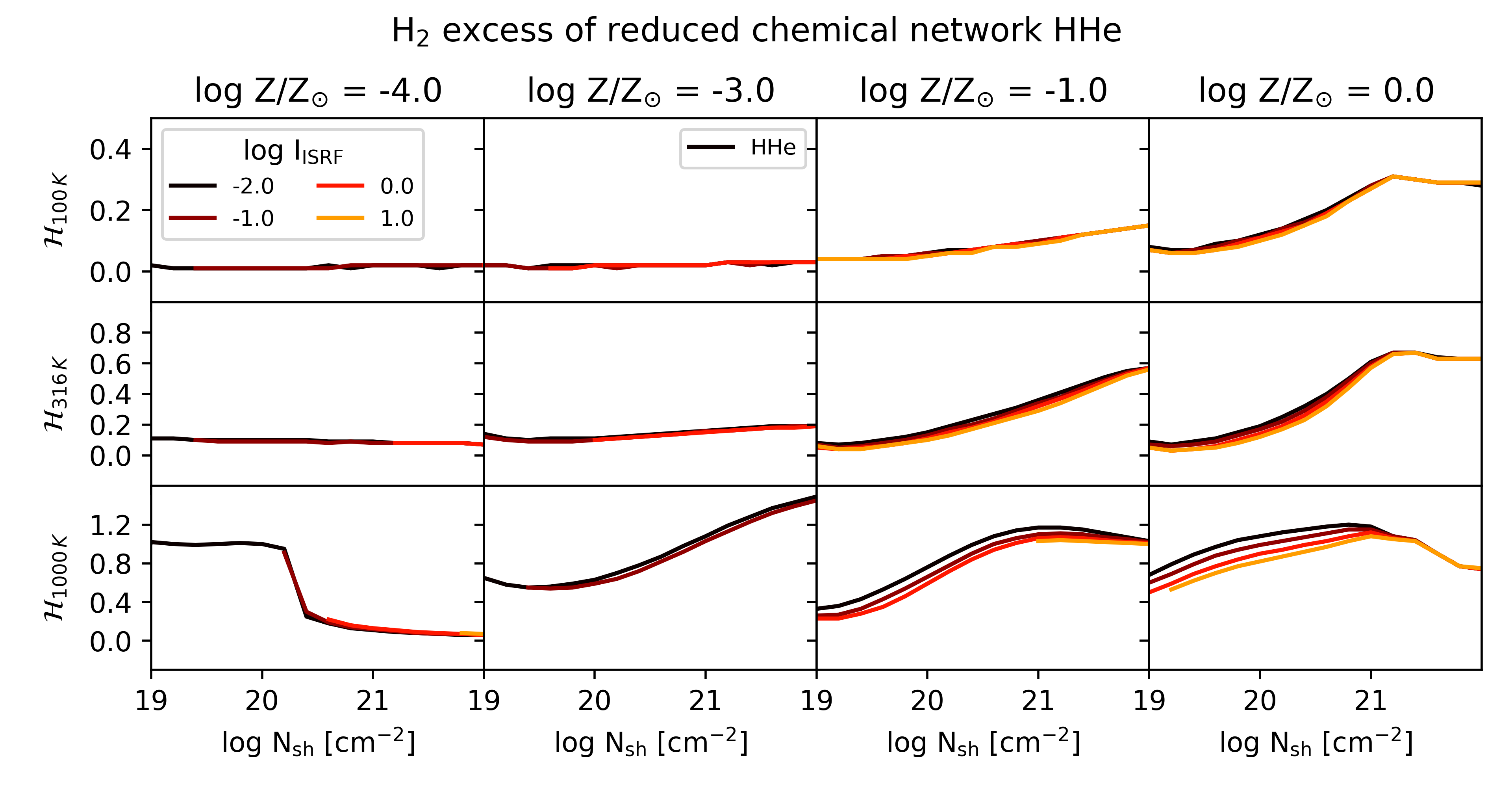}
    \caption{The $\mathrm{H}_2$ excess parameter, $\mathcal{H}$, as defined in equation~(\ref{eq:H}) for the model grid described in section~\ref{sec:modelgrid} between the full chemical network and the chemical network that only includes hydrogen and helium species, \texttt{HHe}. $\mathcal{H}$ is evaluated for different gas temperatures (top row: $\log T  \,[\mathrm{K}]= 2$, middle row: $\log T  \,[\mathrm{K}]= 2.5$, bottom row: $\log T \,[\mathrm{K}]= 3$), gas metallicities (columns from left to right: $\log Z/\mathrm{Z}_{\odot} = -4, -3, -1,0$), and shielding column densities ($\log N_{\mathrm{sh}}$, x-axis). Each panel shows $\mathcal{H} (N_{\mathrm{sh}})$ for different radiation field strengths, from $\log I_{\mathrm{ISRF}} = -2$ to $1$ (from dark to light line colors). Missing lines or line fragments indicate that the \ion{H}{I}-$\mathrm{H}_2$ transition occurs outside the density range of our model grid (see Fig.~\ref{fig:H2excessn}). }
    \label{fig:H2excessH}
\end{figure*}

We use the $\mathrm{H}_2$ excess parameters, $\mathcal{H}$ (equation~\ref{eq:H}), $\log n_{\mathrm{H},\mathcal{E}}$ (equation~\ref{eq:n}), and $\mathcal{E}$ (equation~\ref{eq:E}) to quantify the $\mathrm{H}_2$ excess across different gas temperatures, metallicities, and radiation field strengths. Fig.~\ref{fig:H2excessn} shows the density of the \ion{H}{I} to $\mathrm{H}_2$ transition, for gas temperatures of 100~K ($\log T\,[\mathrm{K}] = 2$, top row), 316~K ($\log T\,[\mathrm{K}] = 2.5$, middle row), and 1000~K ($\log T\,[\mathrm{K}] = 3$, bottom row) and gas metallicities between $\log Z/\mathrm{Z}_{\odot} = -4$ and $\log Z/\mathrm{Z}_{\odot} = 0$ (from left to right columns). The dashed lines are for the reduced chemical network \texttt{HHe} and therefore represent $\log n_{\mathrm{H},\mathcal{E}}$ (equation~\ref{eq:n}), while the solid lines are the transition densities for the full chemistry network (\texttt{Full}). Within each panel, the \ion{H}{I} to $\mathrm{H}_2$ transition shifts to higher densities for stronger radiation fields (line colors from dark: weak radiation field, to bright: strong radiation field). An offset between the transition densities from the reduced (\texttt{HHe}, dashed) and the full (\texttt{Full}, solid) networks is apparent, especially at high metallicities (third and fourth row), but for warm gas (bottom row) persists at low column densities even for metallicities of $\log Z/\mathrm{Z}_{\odot} = -4$ (first column). 

The shift in the transition densities between \texttt{HHe} and \texttt{Full} is quantified by $\mathcal{H}$ (equation~\ref{eq:H}) and shown in Fig.~\ref{fig:H2excessH}. As already seen in Fig.~\ref{fig:H2excessn}, the $\mathrm{H}_2$ excess $\mathcal{H}$  increases with increasing gas temperature (rows; note the different y-axis limits) and is absent for the lowest temperature bin with $\log T\,[\mathrm{K}] = 1.5$ (not shown). We vary the strength of the radiation field by three orders of magnitude, between $\log I_{\mathrm{ISRF}} = -2$ and $1$ (line color) and find that while the \ion{H}{I} to $\mathrm{H}_2$ transition shifts to higher densities for stronger radiation fields (Fig.~\ref{fig:H2excessn}, the relative shift between \texttt{HHe} and \texttt{Full}, $\mathcal{H}$, is largely insensitive to $\log I_{\mathrm{ISRF}}$. Lines are only shown if the transition from \ion{H}{I} to $\mathrm{H}_2$ occurs within the considered density range (Table~\ref{tab:parameters}). For example, for strong radiation fields and low metallicities, this transition is at $\log n_{\mathrm{H}} [\mathrm{cm^{-3}}] > 4$ for lower column densities (see Fig.~\ref{fig:H2excessn}) and therefore not shown. 

\begin{figure*}
    \centering
    \includegraphics[width=\linewidth]{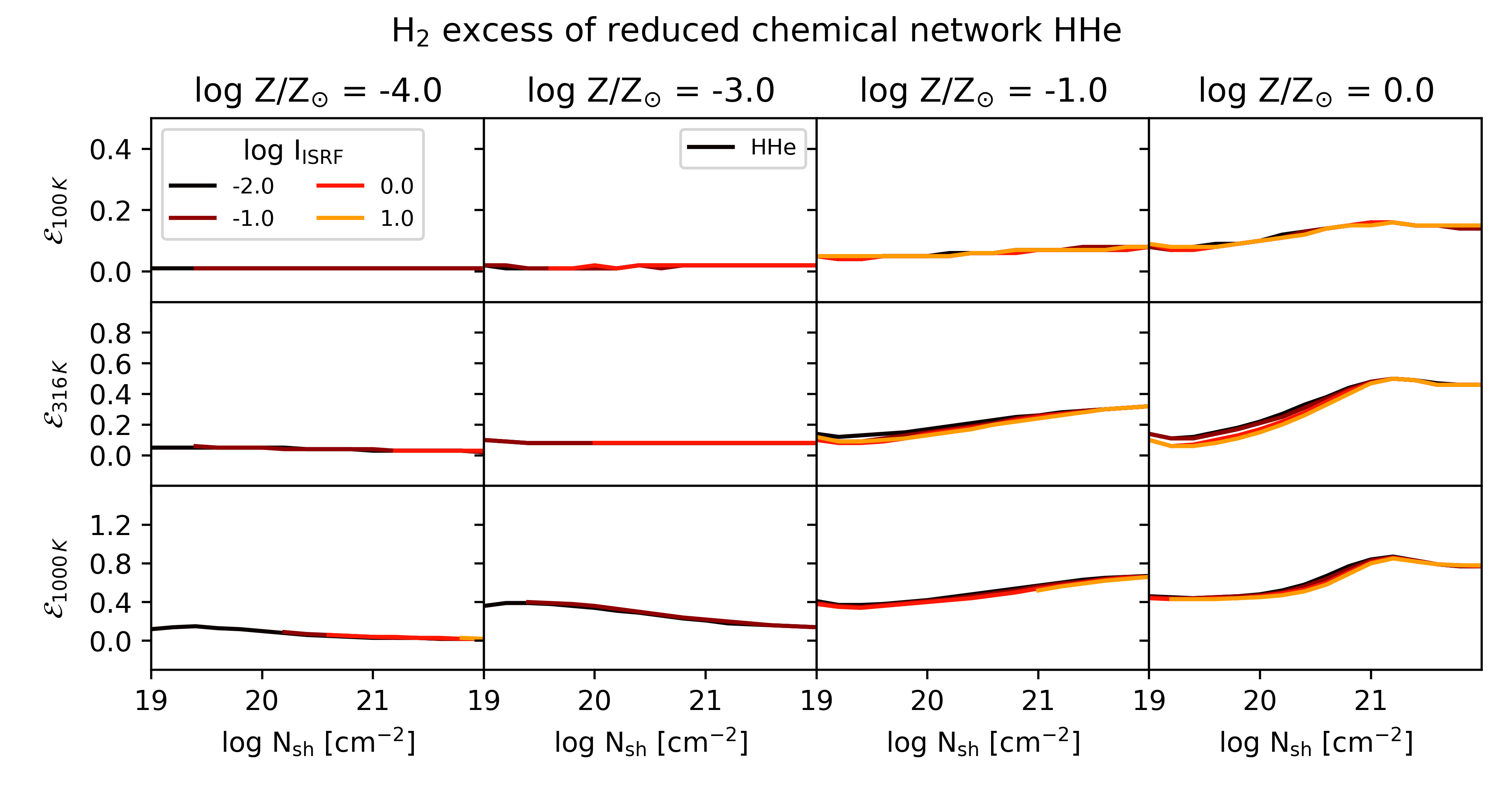}
     \caption{As Fig.~\ref{fig:H2excessH} but for the $\mathrm{H}_2$ excess parameter, $\mathcal{E}$, as defined in equation~(\ref{eq:E}). $\mathcal{H} < \mathcal{E}$ indicates a steep increase in molecular hydrogen fraction, $f_{\mathrm{H2}}$, with density, $n_{\mathrm{H}}$, while $\mathcal{H} > \mathcal{E}$ typically represents a very shallow increase of $f_{\mathrm{H2}}$ with $n_{\mathrm{H}}$ (see Fig.~\ref{fig:H2excessindividual}). }
    \label{fig:H2excessE}
\end{figure*}

The lowest metallicity bins ($\log Z/Z_{\odot} = -4,\,-3$, first and second column) shows a very small ($\mathcal{H}\lesssim 0.2\,\mathrm{dex}$) $\mathrm{H}_2$ excess for low temperatures ($\log T \,[\mathrm{K}] \leq 2.5$, top and middle row). For the highest temperature ($\log T \,[\mathrm{K}] = 3$, bottom row), the $\mathrm{H}_2$ excess is large ($\mathcal{H}_{\mathrm{1000K}} > 0.5\,\mathrm{dex}$, up to $\mathcal{H}_{\mathrm{1000K}} = 1.4\,\mathrm{dex}$ at the highest column density) already at a metallicity of $\log Z/Z_{\odot} = -3$. Even for the lowest metallicity bin, $\log Z/Z_{\odot} = -4$, $\mathcal{H}$ is large ($\approx 1\,\mathrm{dex}$) at column densities of $\log N_{\mathrm{sh}} [\mathrm{cm}^{-2}] \lesssim 20.2$. For these low metallicities, $\mathrm{H}_2$ is the main cooling process at $\log T \,[\mathrm{K}] < 4$ because of the reduced metal-line cooling. A shift in density of the \ion{H}{I}-$\mathrm{H}_2$ transition by between 0.5 and 1.4~dex therefore may affect studies about the cooling, fragmentation, and star formation in the very early stages of galaxy formation. The reason for the $\mathrm{H}_2$ excess at low metallicity is discussed in more detail in section~\ref{sec:lowZ}. 

For low temperatures ($\log T \,[\mathrm{K}] \leq 2.5$), the $\mathrm{H}_2$ excess generally increases with metallicity and is largely independent of the strength of the radiation field ($\log I_{\mathrm{ISRF}} = -2$ to $1$, line color). The largest dependence on the radiation field can be seen in the bottom right panel for solar metallicity ($\log Z/Z_{\odot} = 0$), warm gas ($\log T \,[\mathrm{K}] = 3$), and low column densities ($\log N_{\mathrm{sh}} [\mathrm{cm}^{-2}]\lesssim 21$). Here, the maximum difference in $\mathcal{H}$ is 0.4~dex when varying the radiation field strength by 3~dex. We want to highlight that the normalization of the radiation field together with the cosmic ray rate has a large impact on the density of the \ion{H}{I}-$\mathrm{H}_2$ transition (see Fig.~\ref{fig:H2excessn}), but a small impact on the relative difference in these densities between the full and reduced chemical networks, characterized by $\mathcal{H}$. 

Finally, Fig.~\ref{fig:H2excessE} shows the $\mathrm{H}_2$ excess parameter $\mathcal{E}$ (equation~\ref{eq:E}), representative of the error in $f_{\mathrm{H2}}$ in the \texttt{HHe} network at the \ion{H}{I} to $\mathrm{H}_2$ transition density. While the $\mathrm{H}_2$ excess is still clearly present in $\mathcal{E}$, it appears to be much smaller at low metallicities (first two columns), compared to $\mathcal{H}$ (Fig.~\ref{fig:H2excessH}). This is related to the very shallow increase of $f_{\mathrm{H2}}$ with density for these parameters (see Fig.~\ref{fig:H2excess_single}).

In sections~\ref{sec:lowZ} and \ref{sec:highZ}, we focus on individual calculations from the full model grid for low ($\log Z/Z_{\mathrm{odot}} = -4,\,-3$) and high metallicity ($\log Z/Z_{\mathrm{odot}} = 0$), respectively to investigate the reasons for the $\mathrm{H}_2$ excess in more detail.

\subsection{Reducing the H2 excess by adding metal elements}\label{sec:reducingH2excess}

\begin{figure*}
    \centering
    \includegraphics[width=\linewidth]{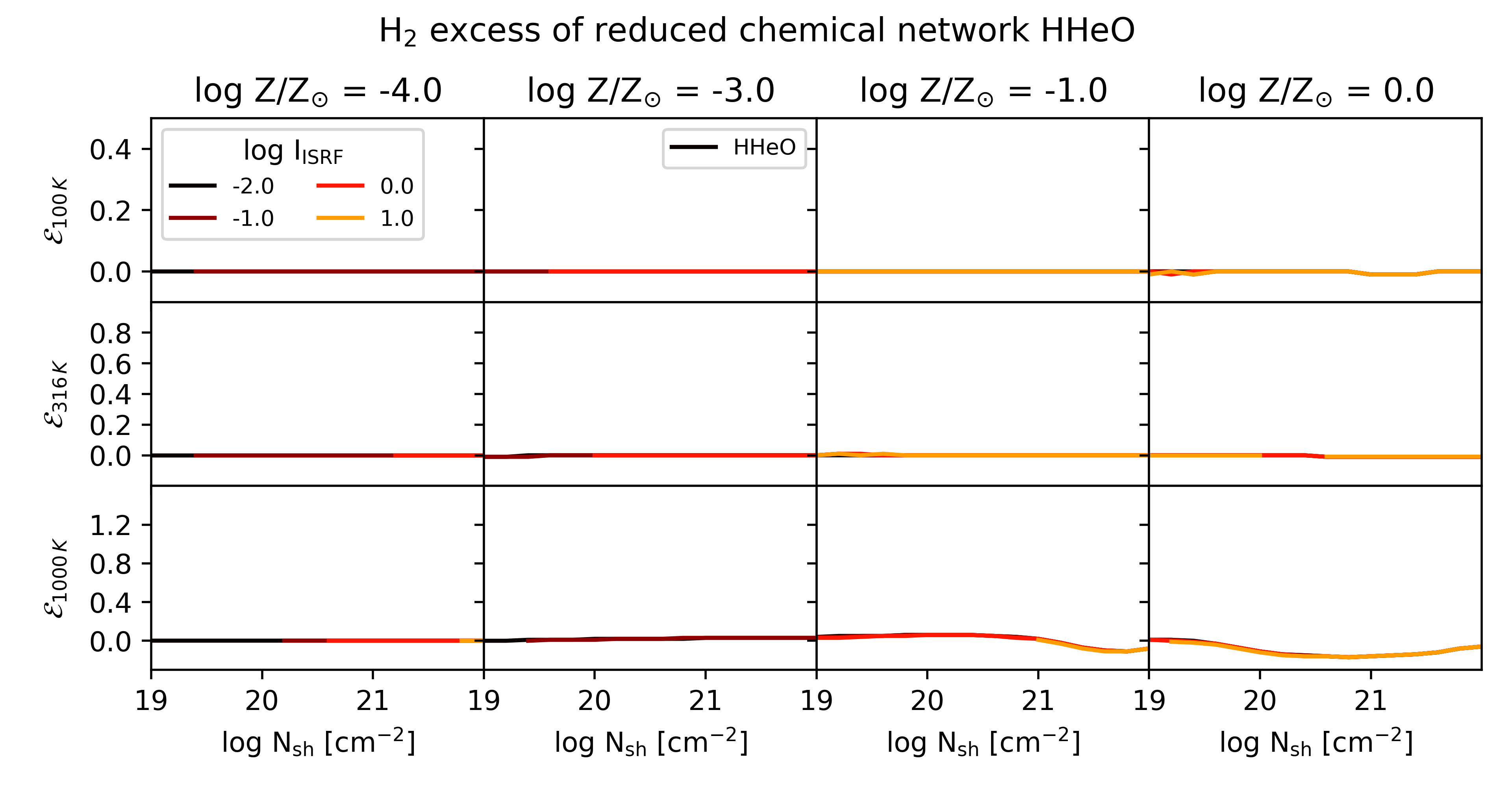}
    \caption{As Fig.~\ref{fig:H2excessE} but for the $\mathrm{H}_2$ excess parameter, $\mathcal{E}$, of a reduced chemical network that includes hydrogen, helium, and oxygen species (\texttt{HHeO}). Reactions with oxygen drastically reduce the $\mathrm{H}_2$ excess, compared to that from the primordial network (compare to Fig.~\ref{fig:H2excessE}).}
    \label{fig:H2excessE_HHeO}
\end{figure*}

In \citet{Ploeckinger2025}, the $\mathrm{H}_2$ fractions from the full network were largely reproduced when adding oxygen to the primoridal chemical network. We repeat this test here by analysing the $\mathrm{H}_2$ excess in \textsc{Cloudy} calculations for a reduced chemical network that includes hydrogen, helium and oxygen species (\texttt{HHeO}, see Table~\ref{tab:chemicalnetworks}). The $\mathrm{H}_2$ excess parameter $\mathcal{E}$ for \texttt{HHeO} is shown in Fig.~\ref{fig:H2excessE_HHeO}. Compared to the results from the primordial network \texttt{HHe} (Fig.~\ref{fig:H2excessE}), the $\mathrm{H}_2$ excess is drastically reduced. Only for metallicities of $\log Z/\mathrm{Z}_{\odot} \geq -1$ and a temperature of $1000\,\mathrm{K}$ (bottom row, two rightmost panels) a small, negative $\mathrm{H}_2$ excess persists which vanishes only when including both carbon and oxygen (\texttt{HHeCO}, not shown).

\begin{figure*}
    \centering
    \includegraphics[width=\linewidth]{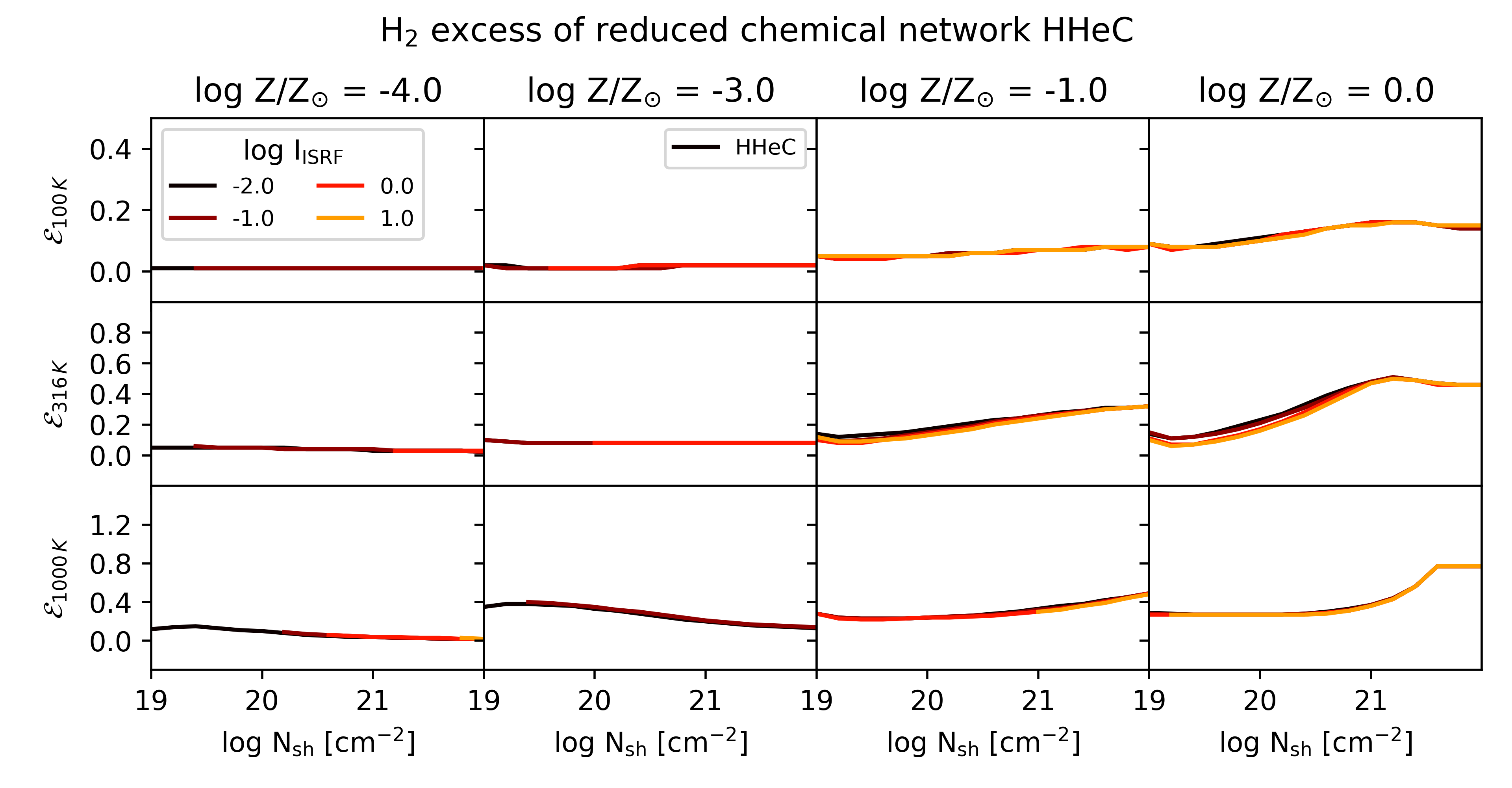}
    \caption{As Fig.~\ref{fig:H2excessE_HHeO} but for a reduced chemical network that includes hydrogen, helium, and carbon species (\texttt{HHeC}). Adding reactions with carbon (but not oxygen) only marginally reduces the $\mathrm{H}_2$ excess from the primordial network (compare to Fig.~\ref{fig:H2excessE}).}
    \label{fig:H2excessE_HHeC}
\end{figure*}

We demonstrate in Fig.~\ref{fig:H2excessE_HHeC} that including carbon instead of oxygen in the chemical network (together with hydrogen and helium) only marginally reduces the $\mathrm{H}_2$ excess. This generalizes the conclusions from \citet{Ploeckinger2025} that reactions with oxygen species are largely responsible for reducing the $\mathrm{H}_2$ fractions in the full chemistry network, relative to the primordial network. The individual reactions between oxygen species and molecular hydrogen in \textsc{Cloudy} are listed in Table~\ref{tab:reactions} and we investigate their individual contributions for both low and high metallicities in the following sections. 

\begin{table}
    \caption{Reactions between oxygen species and molecular hydrogen in \textsc{Cloudy}.}
    \label{tab:reactions}
    \centering
    \begin{tabular}{l@{\hspace{0.5em}}l@{\hspace{0.5em}}l@{\hspace{0.5em}}l@{\hspace{0.5em}}l@{\hspace{1em}}}
    \hline
    O &+ H$_2$ & $\rightarrow$ OH &+ H \\
    O$^+$ &+ H$_2$ & $\rightarrow$ OH$^+$ &+ H   \\ %
    OH &+ H$_2$ & $\rightarrow$ H$_2$O &+ H  \\
    O$_2$ &+ H$_2$ & $\rightarrow$ OH &+ OH  \\
    OH$^+$ &+ H$_2$ & $\rightarrow$ H$_2$O$^+$ &+ H   \\
    H$_2$O$^+$ &+ H$_2$ & $\rightarrow$ H$_3$O$^+$ &+ H  \\
    \hline
\end{tabular}
\end{table}

\subsection{Low metallicity}\label{sec:lowZ}

In order to understand the potential impact on including oxygen for studies on the formation of the first stars in the early Universe, we use a very weak radiation field ($\log I_{\mathrm{ISRF}}=-2$), a low gas metallicity ($\log Z/Z_{\mathrm{\odot}} = -3$), and a temperature of $\log T\,[\mathrm{K}] = 3$, which shows a large $\mathrm{H}_2$ excess (bottom left panel of Fig.~\ref{fig:H2excessH}). For these conditions the $\mathrm{H}_2$ excess parameters for a column density of $\log N_{\mathrm{sh}}\,[\mathrm{cm}^{-2}] = 19$ are $\log n_{\mathrm{H},\mathcal{E}} \,[\mathrm{cm}^{-3}]= 3$, $\mathcal{H} = 0.66 \,\mathrm{dex}$, and $\mathcal{E} = 0.36\,\mathrm{dex}$ (see bottom left panel of Figs.~\ref{fig:H2excessn}, \ref{fig:H2excessH}, and \ref{fig:H2excessE}, respectively). 

For the following analysis, we analyze the destruction rates of $\mathrm{H}_2$ produced with \textsc{Cloudy}'s \verb`save chemistry rates` command which saves the rates at which $\mathrm{H}_2$ is destroyed by species $x$ as

\begin{equation}
    R_x \,[\mathrm{s}^{-1}] = k_{\mathrm{H2},x}\, n_x\quad,
\end{equation}

\noindent
where $k_{\mathrm{H2},x}\,[\mathrm{cm^3\,s}^{-1}]$ is the rate coefficient of the reaction between $\mathrm{H}_2$ and species $x$, and $n_x \, [\mathrm{cm}^{-3}]$ is the number density of species $x$. For the oxygen species that react with $\mathrm{H}_2$, the rate coefficients for the neutral species, $\mathrm{O}$, $\mathrm{OH}$, and $\mathrm{O}_2$, have a very steep temperature dependence, while the rate coefficients for the ionized species, $\mathrm{O}^+$, $\mathrm{OH}^+$, and $\mathrm{H}_2\mathrm{O}^+$, are constant (see e.g. table B1 in \citealp{Glover2010}). At constant temperature, the reaction rates, $R_x$, are therefore always proportional to $n_{x}$. In turn, at higher temperatures (e.g. $\log T \,[K] = 3$), neutral species may contribute more to the destruction of $\mathrm{H}_2$ than at lower temperatures (e.g. $\log T \,[K] = 2$), if the decrease in $n_{x}$ is counterbalanced by the drastic increase in $k_{\mathrm{H2},x}$.

\begin{figure}
    \centering
    \includegraphics[width=\linewidth]{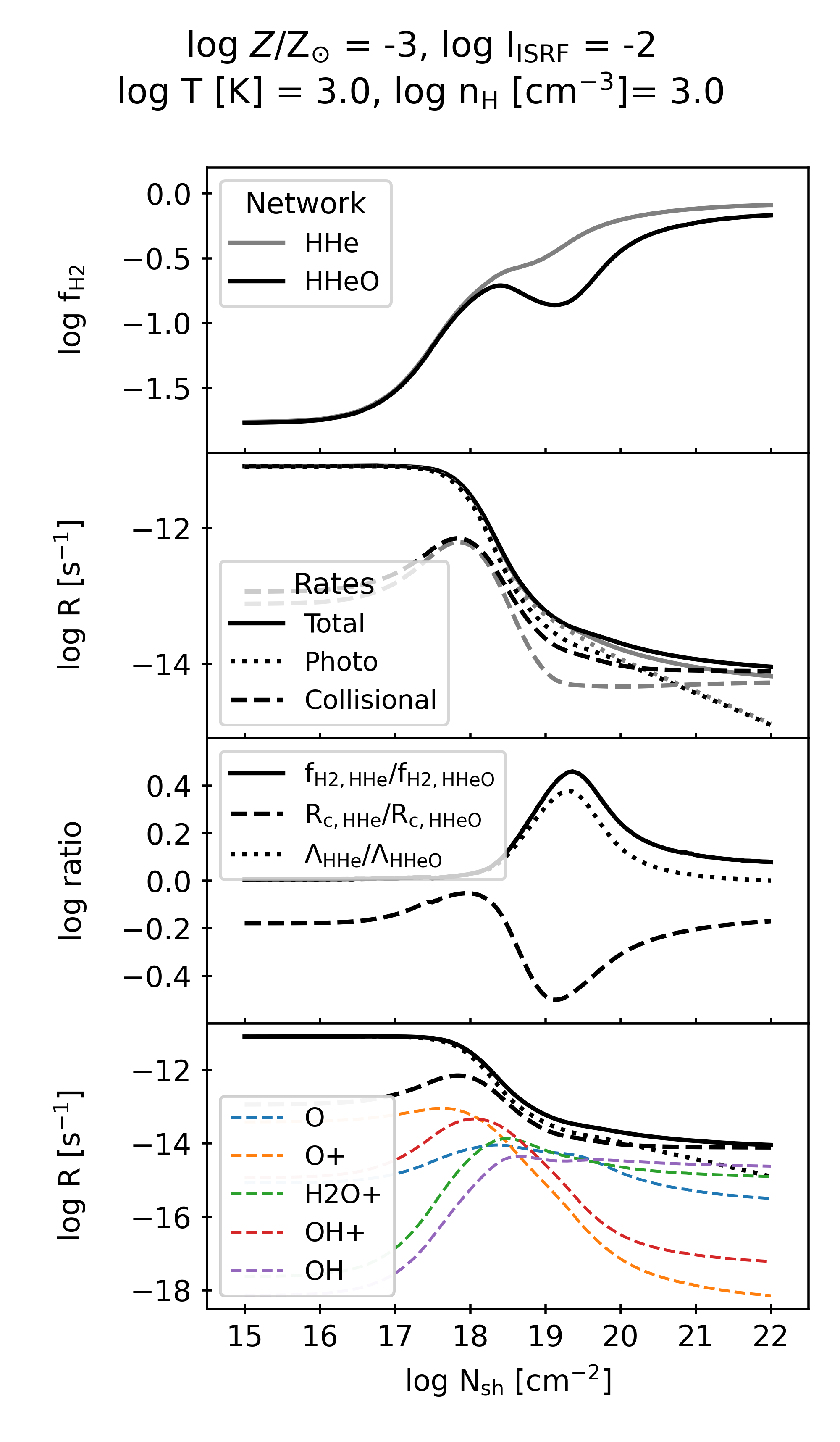}
    \caption{In-depth analysis of one individual low-metallicity \textsc{Cloudy} grid point. The modeled slab of gas has a metallicity of $\log Z/\mathrm{Z}_{\odot} = -3$, a temperature of $\log T\,[\mathrm{K}] = 3$, a density of $\log n_{\mathrm{H}}\,[\mathrm{cm}^{-3}] = 3$ and is illuminated by a weak interstellar radiation field of $\log I_{\mathrm{ISRF}} = \log G = -2$. We show the $\mathrm{H}_2$ fractions, $f_{\mathrm{H2}}$ (first panel), at an increasing depth (expressed as shielding column density, $\log N_{\mathrm{sh}}$) into the gas cloud, for the chemical networks \texttt{HHe} (grey lines) and \texttt{HHeO} (black lines). The total reaction rates for processes that destroy $\mathrm{H}_2$ can be found in the second panel (solid lines), together with the contributions from reactions that involve photons (dotted lines) and collisions (dashed lines). The third panel compares the $\mathrm{H}_2$ fractions (solid line), the reaction rates from collisions (dashed line), and the total cooling rates (dotted line) from the chemical networks with and without oxygen. The fourth panel repeats the $\mathrm{H}_2$ destruction rates from the \texttt{HHeO} network (black lines, as in second panel), while the colored dashed lines show the rates of individual reaction of $\mathrm{H}_2$ with various oxygen species (see legend and Table~\ref{tab:reactions}).}
    \label{fig:destruction_lowZ_logNsh19}
\end{figure}

The top panel of Fig.~\ref{fig:destruction_lowZ_logNsh19} shows the molecular hydrogen fraction, $f_{\mathrm{H2}}$, for this individual \textsc{Cloudy} run ($\log n_{\mathrm{H}} = \log n_{\mathrm{H},\mathcal{E}}$) using the chemical network with (\texttt{HHeO}, black line) and without (\texttt{HHe}, grey line) oxygen. At low shielding column densities ($\log N_{\mathrm{sh}} \,[\mathrm{cm}^{-2}]\lesssim 18$), the molecular hydrogen fractions from the two networks are identical but they disagree at higher column densities. The second panel shows the destruction rates of $\mathrm{H}_2$ per $\mathrm{H}_2$ molecule by photons (photo-dissociation, $\gamma + \mathrm{H}_2\rightarrow \mathrm{H}+\mathrm{H}$, plus photo-ionization, $\gamma + \mathrm{H}_2\rightarrow \mathrm{H}_2^{+}+\mathrm{e}^{-}$, dotted lines) and by collisional processes (dashed lines), as well as the total destruction rates (solid lines) for both chemical networks (grey and black, as in top panel). Here we see that the destruction of $\mathrm{H}_2$ is dominated by photo-dissociation and photo-ionization in both chemical networks for $\log N_{\mathrm{sh}} \,[\mathrm{cm}^{-2}]\lesssim 18$, which explains the insensitivity of $f_{\mathrm{H2}}$ on the inclusion of oxygen in the chemical network. At higher column densities, collisional dissociation processes become important and their rates exceed those for reactions with photons for $\log N_{\mathrm{sh}} \,[\mathrm{cm}^{-2}]\gtrsim 20$ for the network with oxygen. 

At a column density of $\log N_{\mathrm{sh}} \,[\mathrm{cm}^{-2}] = 19$, at which $f_{\mathrm{H2}}$ is much higher in the network without oxygen than in the network with oxygen ($\mathcal{E} = 0.36\,\mathrm{dex}$), the total $\mathrm{H}_2$ destruction rates also show the largest discrepancy between the two networks. This is shown more quantitatively in the third panel of Fig.~\ref{fig:destruction_lowZ_logNsh19}. Here, the black solid line shows the ratio of $f_{\mathrm{H2}}$ of both networks (i.e. the ratio of the lines in the top panel) and the dashed line shows the ratio of the collisional destruction rates of $\mathrm{H}_2$ (i.e. the ratio the dashed lines in the second panel). We conclude that the overestimation of $f_{\mathrm{H2}}$ directly correlates with the underestimated $\mathrm{H}_2$ destruction rates in the network without oxygen for column densities at which collisional dissociation processes are important. In this example, the total radiative cooling is dominated by cooling from $\mathrm{H}_2$ ($\gtrsim 99$ per cent of the total cooling rate) and the difference in $f_{\mathrm{H2}}$ directly translates into comparable differences it the cooling rates (dotted line, third panel). 

The coloured lines in the fourth panel of Fig.~\ref{fig:destruction_lowZ_logNsh19} provide additional insight into the reaction rates involving oxygen species that are responsible for an increased $\mathrm{H}_2$ destruction rate and therefore the decreased molecular fraction $f_{\mathrm{H2}}$ in the \texttt{HHeO} network. At low shielding column densities ($\log N_{\mathrm{sh}} \,[\mathrm{cm}^{-2}]\lesssim 18$), the reactions with $\mathrm{O}^+$ increases the total collisional $\mathrm{H}_2$ destruction rates but this barely affects $f_{\mathrm{H2}}$ because the dominant $\mathrm{H}_2$ destruction processes are photo-dissociation and photo-ionization. At $\log N_{\mathrm{sh}} \,[\mathrm{cm}^{-2}] = 19$, close to the maximal discrepancy between the two networks, reactions of $\mathrm{H}_2$ with $\mathrm{O}$, $\mathrm{H}_2\mathrm{O}^{+}$, and $\mathrm{OH}$ together increase the collisional destruction rates of $\mathrm{H}_2$ (compare dashed lines in second panel). Reactions of $\mathrm{H}_2$ with $\mathrm{OH}^+$ and $\mathrm{O}^{+}$ further contribute to the destruction of $\mathrm{H}_2$ at $\log N_{\mathrm{sh}} \,[\mathrm{cm}^{-2}] = 19$ but decrease steeply with increasing column density. 

At the largest column densities, $\log N_{\mathrm{sh}} [\mathrm{cm}^{-3}]= 22$, photo-ionization and photo-dissociation of $\mathrm{H}_2$ is negligible. Here, the majority of oxygen is in the form of neutral $\mathrm{H}_2\mathrm{O}$ which does not react with $\mathrm{H}_2$, which explains the better agreement of $f_{\mathrm{H2}}$ between \texttt{HHe} and \texttt{HHeO}.

\subsection{High metallicity}\label{sec:highZ}

\begin{figure}
    \centering
    \includegraphics[width=\linewidth]{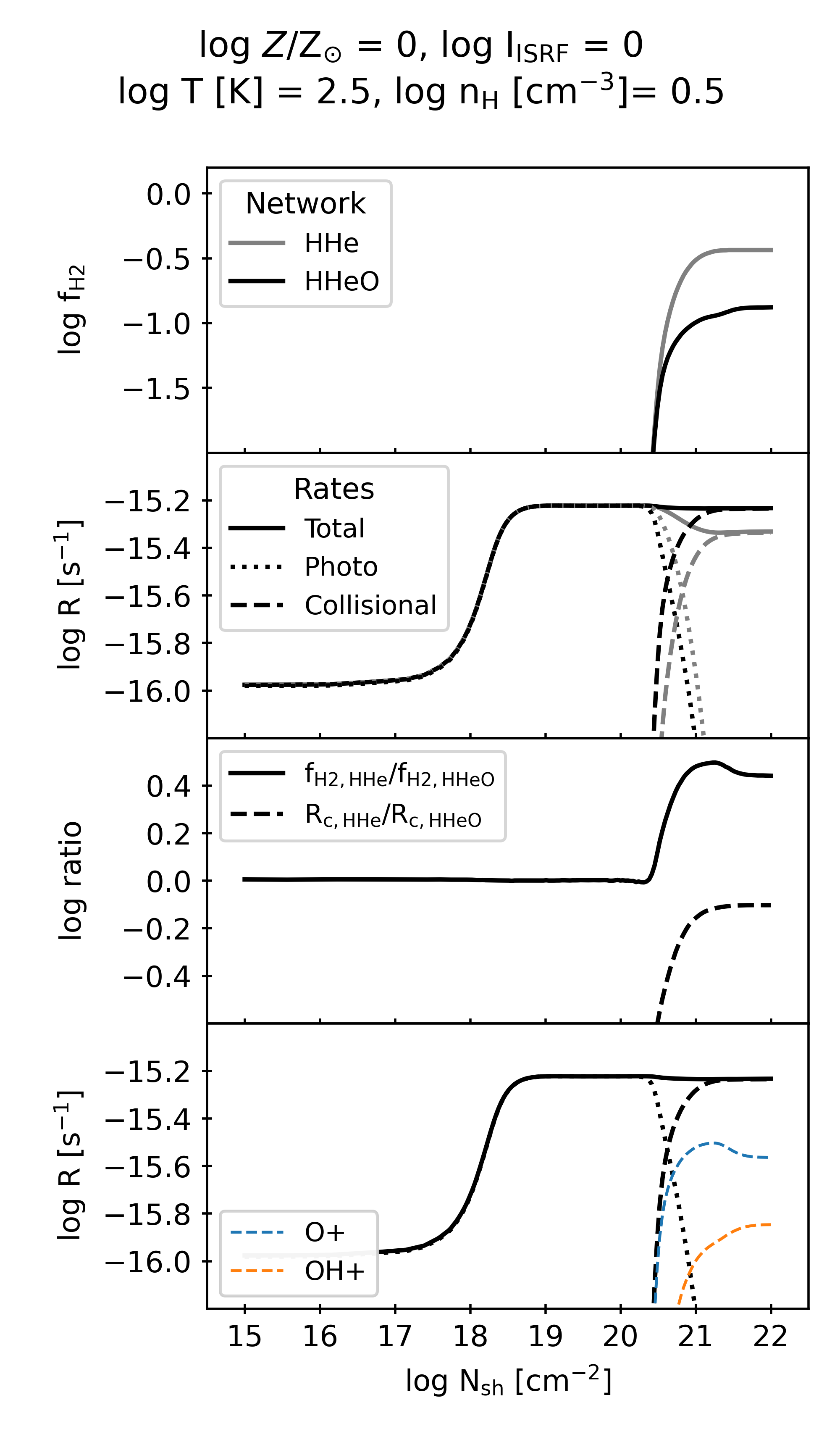}
    \caption{As Fig.~\ref{fig:destruction_lowZ_logNsh19} but for gas properties typical of the \ion{H}{I}-$\mathrm{H}_2$ transition in the ISM of Milky Way mass galaxies: $\log I_{\mathrm{ISRF}} = 0$, $\log Z/Z_{\odot} = 0$, $\log T\,[\mathrm{K}] = 2.5$, $\log f_{\mathrm{H2}}=-0.5$. We omit the ratio of the cooling rates in the third panel because here cooling is dominated by metal-line cooling, which (by design) is absent in \texttt{HHe}. }
    \label{fig:destruction_highZ_logNsh21}
\end{figure}

We repeat the analysis from section~\ref{sec:lowZ} for conditions typical for the interstellar medium of the solar neighborhood ($\log I_{\mathrm{ISRF}} = 0$, $\log Z/Z_{\odot} = 0$). For a temperature of $\log T\,[\mathrm{K}] = 2.5$ and a shielding column density of $\log N_{\mathrm{sh}}\,[\mathrm{cm}^{-2}]= 21$, the density of the \ion{H}{I} to $\mathrm{H}_2$ transition is at a density of $\log n_{\mathrm{H}} \,[\mathrm{cm}^{-3}] = 0.5$ in the network \texttt{HHe}. Here, the $\mathrm{H}_2$ excess is $\mathcal{H} = 0.60\,\mathrm{dex}$ and $\mathcal{E} = 0.48\,\mathrm{dex}$ (see right middle panels of Figs.~\ref{fig:H2excessH} and \ref{fig:H2excessE}). 

The results are shown in Fig.~\ref{fig:destruction_highZ_logNsh21}. As for the low metallicity case, $f_{\mathrm{H2}}$ is the same in both networks for column densities at which $\mathrm{H}_2$ is mainly destroyed by photo-dissociation and photo-ionization (see solid line in third panel). At column densities of $\log N_{\mathrm{sh}}\,[\mathrm{cm}^{-2}]\gtrsim 20.5$, collisional dissociation dominates the destruction of $\mathrm{H}_2$ and the slightly elevated destruction rate when including oxygen species (compare dashed lines in second panel or see dashed line in third panel) results in a persistent decrease in $f_{\mathrm{H2}}$ (compare lines in first panel or see solid line in third panel). The total collisional destruction rate of $\mathrm{H}_2$ is mainly increased by reactions with $\mathrm{O}^{+}$ and $\mathrm{OH}^{+}$ at high column densities. 

\section{Summary}\label{sec:summary}

We have shown that the increased $\mathrm{H}_2$ fraction (here: $\mathrm{H}_2$ excess) in primordial chemical networks found in \citet{Ploeckinger2025} using the \textsc{Chimes} chemical network is not limited to specific choices for the radiation fields and shielding column densities. We used \textsc{Cloudy} version 25.00 and found systematic differences in the densities of the \ion{H}{I}-$\mathrm{H}_2$ transition between primordial and full chemical networks, even if both are supposed to represent the same gas-phase metallicity.   

We have shown in Section~\ref{sec:results} that an $\mathrm{H}_2$ excess is present over a large range of parameters, increasing both with gas metallicity and gas temperature, while being largely insensitive to the strength of the radiation field. The typical temperature of molecular gas in the ISM is a few tens of K and for these temperatures, we did not find an $\mathrm{H}_2$ excess. Here, oxygen species, including oxygen-bearing molecules, are largely neutral and either do not react with $\mathrm{H}_2$ (such as $\mathrm{H}_2\mathrm{O}$) or reactions are inefficient because of the steep temperature dependence of their rate coefficients (such as for $\mathrm{O}$, $\mathrm{OH}$, and $\mathrm{O}_2$). Studies of very cold gas ($T\lesssim 50\,\mathrm{K}$) are therefore unaffected by the $\mathrm{H}_2$ excess. 

On the other hand, for gas that is cooling from higher temperatures ($\log T\,[\mathrm{K}]\gtrsim 2.5\,\mathrm{K}$), the transition from atomic to molecular gas may be shifted to lower densities by up to an order of magnitude, when neglecting reactions with oxygen. This is particularly important for gas with very low metallicity ($\log Z/\mathrm{Z}_{\odot}\lesssim -3$), for which the total radiative cooling rate at $T\lesssim10^4\,\mathrm{K}$ is dominated by $\mathrm{H}_2$ cooling. Here, the $\mathrm{H}_2$ excess directly leads to a cooling rate excess which may affect the cooling and fragmentation of gas in very early Universe.

We note that the molecular hydrogen fraction, $f_{\mathrm{H2}}$, is sensitive to the shielding column density, the assumed radiation field (for low shielding column densities) and cosmic ray rates (for high shielding column densities), all of which are often not very accurately known in simulations. Variations in either of these parameters may have a larger effect on $f_{\mathrm{H2}}$ than the inclusion of oxygen. Nonetheless, it is important to recognize this systematic $\mathrm{H}_2$ excess. Especially in chemical networks that already track oxygen species, adding individual reactions with $\mathrm{H}_2$ may not increase the computational cost. 

In applications for which including reactions with oxygen are prohibitively expensive, it may be worth estimating if the magnitude of the $\mathrm{H}_2$ excess is larger or smaller than the expected non-equilibrium effects. Pre-tabulated $\mathrm{H}_2$ fractions, calculated with a full chemical network under the assumption of chemical equilibrium, may be more accurate than expensive non-equilibrium calculations that do not include oxgyen reactions. 

In a forthcoming paper, we will present results from cosmological simulations within the \textsc{Colibre} project \citep{Chaikin2025a, Schaye2025colibre} that use \textsc{Hybrid-Chimes} in both \texttt{HHe} or \texttt{HHeO} mode and discuss the $\mathrm{H}_2$ excess in simulated galaxies.

\section*{Acknowledgments}

I would like to thank Alvaro Hacar and Moa Huppenkothen for insightful discussions and the anonymous referee for a constructive report that helped clarify and improve the manuscript. 
This research was funded by the Austrian Science Fund (FWF) through grant-DOI: 10.55776/V982. The computational results have been achieved using the Austrian Scientific Computing (ASC) infrastructure.

\section*{Data Availability}

\textsc{Cloudy} is a public code, available from \href{www.nublado.org}{www.nublado.org} and we use version 25.00 in this work. For full transparency and reproducibility, the scripts to set up the \textsc{Cloudy} grids as well as the scripts that produce the figures in this paper are publicly available from \href{https://gitlab.phaidra.org/h2excess/h2excess_analysis}{gitlab.phaidra.org/h2excess}.



\bibliographystyle{mnras}
\bibliography{molecularhydrogenexcess} 

@ARTICLE{Mathis1983,
       author = {{Mathis}, J.~S. and {Mezger}, P.~G. and {Panagia}, N.},
        title = "{Interstellar radiation field and dust temperatures in the diffuse interstellar medium and in giant molecular clouds}",
      journal = {\aap},
         year = 1983,
        month = nov,
       volume = {128},
        pages = {212-229},
       adsurl = {https://ui.adsabs.harvard.edu/abs/1983A&A...128..212M}
}

@ARTICLE{Bregman1986,
       author = {{Bregman}, Joel N. and {Harrington}, J. Patrick},
        title = "{Photoionization in the Halo of the Galaxy}",
      journal = {\apj},
         year = 1986,
        month = oct,
       volume = {309},
        pages = {833},
          doi = {10.1086/164652},
       adsurl = {https://ui.adsabs.harvard.edu/abs/1986ApJ...309..833B}
}

@INPROCEEDINGS{Black1987,
       author = {{Black}, John H.},
        title = "{Heating and Cooling of the Interstellar Gas}",
    booktitle = {Interstellar Processes},
         year = 1987,
       editor = {{Hollenbach}, David J. and {Thronson}, Harley A., Jr.},
      series = "{Astrophysics and Space Science Library}",
       volume = {134},
        month = jan,
        pages = {731},
          doi = {10.1007/978-94-009-3861-8_27},
       adsurl = {https://ui.adsabs.harvard.edu/abs/1987ASSL..134..731B}
}

@ARTICLE{Indriolo2015,
       author = {{Indriolo}, Nick and {Neufeld}, D.~A. and {Gerin}, M. and {Schilke}, P. and {Benz}, A.~O. and {Winkel}, B. and {Menten}, K.~M. and {Chambers}, E.~T. and {Black}, John H. and {Bruderer}, S. and {Falgarone}, E. and {Godard}, B. and {Goicoechea}, J.~R. and {Gupta}, H. and {Lis}, D.~C. and {Ossenkopf}, V. and {Persson}, C.~M. and {Sonnentrucker}, P. and {van der Tak}, F.~F.~S. and {van Dishoeck}, E.~F. and {Wolfire}, Mark G. and {Wyrowski}, F.},
        title = "{Herschel Survey of Galactic OH$^{+}$, H$_{2}$O$^{+}$, and H$_{3}$O$^{+}$: Probing the Molecular Hydrogen Fraction and Cosmic-Ray Ionization Rate}",
      journal = {\apj},
         year = 2015,
        month = feb,
       volume = {800},
       number = {1},
          eid = {40},
        pages = {40},
          doi = {10.1088/0004-637X/800/1/40},
archivePrefix = {arXiv},
       eprint = {1412.1106},
 primaryClass = {astro-ph.GA},
       adsurl = {https://ui.adsabs.harvard.edu/abs/2015ApJ...800...40I}
}

@ARTICLE{chimes2014optthin,
       author = {{Richings}, A.~J. and {Schaye}, J. and {Oppenheimer}, B.~D.},
        title = "{Non-equilibrium chemistry and cooling in the diffuse interstellar medium - I. Optically thin regime}",
      journal = {\mnras},
         year = 2014,
        month = jun,
       volume = {440},
       number = {4},
        pages = {3349-3369},
          doi = {10.1093/mnras/stu525},
archivePrefix = {arXiv},
       eprint = {1401.4719},
 primaryClass = {astro-ph.GA},
       adsurl = {https://ui.adsabs.harvard.edu/abs/2014MNRAS.440.3349R}
}

@ARTICLE{chimes2014shielded,
       author = {{Richings}, A.~J. and {Schaye}, J. and {Oppenheimer}, B.~D.},
        title = "{Non-equilibrium chemistry and cooling in the diffuse interstellar medium - II. Shielded gas}",
      journal = {\mnras},
         year = 2014,
        month = aug,
       volume = {442},
       number = {3},
        pages = {2780-2796},
          doi = {10.1093/mnras/stu1046},
archivePrefix = {arXiv},
       eprint = {1403.6155},
 primaryClass = {astro-ph.GA},
       adsurl = {https://ui.adsabs.harvard.edu/abs/2014MNRAS.442.2780R}
}

@ARTICLE{Asplund2009,
       author = {{Asplund}, Martin and {Grevesse}, Nicolas and {Sauval}, A. Jacques and {Scott}, Pat},
        title = "{The Chemical Composition of the Sun}",
      journal = {\araa},
         year = 2009,
        month = sep,
       volume = {47},
       number = {1},
        pages = {481-522},
          doi = {10.1146/annurev.astro.46.060407.145222},
archivePrefix = {arXiv},
       eprint = {0909.0948},
 primaryClass = {astro-ph.SR},
       adsurl = {https://ui.adsabs.harvard.edu/abs/2009ARA&A..47..481A}
}

@ARTICLE{Jenkins2009,
       author = {{Jenkins}, Edward B.},
        title = "{A Unified Representation of Gas-Phase Element Depletions in the Interstellar Medium}",
      journal = {\apj},
         year = 2009,
        month = aug,
       volume = {700},
       number = {2},
        pages = {1299-1348},
          doi = {10.1088/0004-637X/700/2/1299},
archivePrefix = {arXiv},
       eprint = {0905.3173},
 primaryClass = {astro-ph.GA},
       adsurl = {https://ui.adsabs.harvard.edu/abs/2009ApJ...700.1299J}
}

@ARTICLE{Habing1968,
       author = {{Habing}, H.~J.},
        title = "{The interstellar radiation density between 912 A and 2400 A}",
      journal = {\bain},
         year = 1968,
        month = jan,
       volume = {19},
        pages = {421},
       adsurl = {https://ui.adsabs.harvard.edu/abs/1968BAN....19..421H}
}

@ARTICLE{swift2024,
       author = {{Schaller}, Matthieu and {Borrow}, Josh and {Draper}, Peter W. and {Ivkovic}, Mladen and {McAlpine}, Stuart and {Vandenbroucke}, Bert and {Bah{\'e}}, Yannick and {Chaikin}, Evgenii and {Chalk}, Aidan B.~G. and {Chan}, Tsang Keung and {Correa}, Camila and {van Daalen}, Marcel and {Elbers}, Willem and {Gonnet}, Pedro and {Hausammann}, Lo{\"\i}c and {Helly}, John and {Hu{\v{s}}ko}, Filip and {Kegerreis}, Jacob A. and {Nobels}, Folkert S.~J. and {Ploeckinger}, Sylvia and {Revaz}, Yves and {Roper}, William J. and {Ruiz-Bonilla}, Sergio and {Sandnes}, Thomas D. and {Uyttenhove}, Yolan and {Willis}, James S. and {Xiang}, Zhen},
        title = "{SWIFT: A modern highly-parallel gravity and smoothed particle hydrodynamics solver for astrophysical and cosmological applications}",
      journal = {\mnras},
         year = 2024,
        month = may,
       volume = {530},
       number = {2},
        pages = {2378-2419},
          doi = {10.1093/mnras/stae922},
archivePrefix = {arXiv},
       eprint = {2305.13380},
 primaryClass = {astro-ph.IM},
       adsurl = {https://ui.adsabs.harvard.edu/abs/2024MNRAS.530.2378S}
}

@ARTICLE{Glover2010,
       author = {{Glover}, S.~C.~O. and {Federrath}, C. and {Mac Low}, M. -M. and {Klessen}, R.~S.},
        title = "{Modelling CO formation in the turbulent interstellar medium}",
      journal = {\mnras},
         year = 2010,
        month = may,
       volume = {404},
       number = {1},
        pages = {2-29},
          doi = {10.1111/j.1365-2966.2009.15718.x},
archivePrefix = {arXiv},
       eprint = {0907.4081},
 primaryClass = {astro-ph.SR},
       adsurl = {https://ui.adsabs.harvard.edu/abs/2010MNRAS.404....2G}
}

@ARTICLE{Wakelam2017,
       author = {{Wakelam}, Valentine and {Bron}, Emeric and {Cazaux}, Stephanie and {Dulieu}, Francois and {Gry}, C{\'e}cile and {Guillard}, Pierre and {Habart}, Emilie and {Hornek{\ae}r}, Liv and {Morisset}, Sabine and {Nyman}, Gunnar and {Pirronello}, Valerio and {Price}, Stephen D. and {Valdivia}, Valeska and {Vidali}, Gianfranco and {Watanabe}, Naoki},
        title = "{H$_{2}$ formation on interstellar dust grains: The viewpoints of theory, experiments, models and observations}",
      journal = {Molecular Astrophysics},
         year = 2017,
        month = dec,
       volume = {9},
        pages = {1-36},
          doi = {10.1016/j.molap.2017.11.001},
archivePrefix = {arXiv},
       eprint = {1711.10568},
 primaryClass = {astro-ph.GA},
       adsurl = {https://ui.adsabs.harvard.edu/abs/2017MolAs...9....1W}
}

@ARTICLE{Grackle2017,
       author = {{Smith}, Britton D. and {Bryan}, Greg L. and {Glover}, Simon C.~O. and {Goldbaum}, Nathan J. and {Turk}, Matthew J. and {Regan}, John and {Wise}, John H. and {Schive}, Hsi-Yu and {Abel}, Tom and {Emerick}, Andrew and {O'Shea}, Brian W. and {Anninos}, Peter and {Hummels}, Cameron B. and {Khochfar}, Sadegh},
        title = "{GRACKLE: a chemistry and cooling library for astrophysics}",
      journal = {\mnras},
         year = 2017,
        month = apr,
       volume = {466},
       number = {2},
        pages = {2217-2234},
          doi = {10.1093/mnras/stw3291},
archivePrefix = {arXiv},
       eprint = {1610.09591},
 primaryClass = {astro-ph.CO},
       adsurl = {https://ui.adsabs.harvard.edu/abs/2017MNRAS.466.2217S}
}

@ARTICLE{Glassgold1973,
       author = {{Glassgold}, A.~E. and {Langer}, William D.},
        title = "{Cosmic-Ray Heating and Molecular Cooling of Dense Clouds}",
      journal = {\apjl},
         year = 1973,
        month = feb,
       volume = {179},
        pages = {L147},
          doi = {10.1086/181137},
       adsurl = {https://ui.adsabs.harvard.edu/abs/1973ApJ...179L.147G}
}

@ARTICLE{Hyacinth2024,
       author = {{Khatri}, Prachi and {Porciani}, Cristiano and {Romano-D{\'\i}az}, Emilio and {Seifried}, Daniel and {Sch{\"a}be}, Alexander},
        title = "{HYACINTH: HYdrogen And Carbon chemistry in the INTerstellar medium in Hydro simulations}",
      journal = {\aap},
         year = 2024,
        month = aug,
       volume = {688},
          eid = {A194},
        pages = {A194},
          doi = {10.1051/0004-6361/202449640},
archivePrefix = {arXiv},
       eprint = {2402.11023},
 primaryClass = {astro-ph.GA},
       adsurl = {https://ui.adsabs.harvard.edu/abs/2024A&A...688A.194K}
}

@ARTICLE{Agora2014,
       author = {{Kim}, Ji-hoon and {Abel}, Tom and {Agertz}, Oscar and {Bryan}, Greg L. and {Ceverino}, Daniel and {Christensen}, Charlotte and {Conroy}, Charlie and {Dekel}, Avishai and {Gnedin}, Nickolay Y. and {Goldbaum}, Nathan J. and {Guedes}, Javiera and {Hahn}, Oliver and {Hobbs}, Alexander and {Hopkins}, Philip F. and {Hummels}, Cameron B. and {Iannuzzi}, Francesca and {Keres}, Dusan and {Klypin}, Anatoly and {Kravtsov}, Andrey V. and {Krumholz}, Mark R. and {Kuhlen}, Michael and {Leitner}, Samuel N. and {Madau}, Piero and {Mayer}, Lucio and {Moody}, Christopher E. and {Nagamine}, Kentaro and {Norman}, Michael L. and {Onorbe}, Jose and {O'Shea}, Brian W. and {Pillepich}, Annalisa and {Primack}, Joel R. and {Quinn}, Thomas and {Read}, Justin I. and {Robertson}, Brant E. and {Rocha}, Miguel and {Rudd}, Douglas H. and {Shen}, Sijing and {Smith}, Britton D. and {Szalay}, Alexander S. and {Teyssier}, Romain and {Thompson}, Robert and {Todoroki}, Keita and {Turk}, Matthew J. and {Wadsley}, James W. and {Wise}, John H. and {Zolotov}, Adi and {AGORA Collaboration29},the},
        title = "{The AGORA High-resolution Galaxy Simulations Comparison Project}",
      journal = {\apjs},
         year = 2014,
        month = jan,
       volume = {210},
       number = {1},
          eid = {14},
        pages = {14},
          doi = {10.1088/0067-0049/210/1/14},
archivePrefix = {arXiv},
       eprint = {1308.2669},
 primaryClass = {astro-ph.GA},
       adsurl = {https://ui.adsabs.harvard.edu/abs/2014ApJS..210...14K}
}

@ARTICLE{Dave2019,
       author = {{Dav{\'e}}, Romeel and {Angl{\'e}s-Alc{\'a}zar}, Daniel and {Narayanan}, Desika and {Li}, Qi and {Rafieferantsoa}, Mika H. and {Appleby}, Sarah},
        title = "{SIMBA: Cosmological simulations with black hole growth and feedback}",
      journal = {\mnras},
         year = 2019,
        month = jun,
       volume = {486},
       number = {2},
        pages = {2827-2849},
          doi = {10.1093/mnras/stz937},
archivePrefix = {arXiv},
       eprint = {1901.10203},
 primaryClass = {astro-ph.GA},
       adsurl = {https://ui.adsabs.harvard.edu/abs/2019MNRAS.486.2827D}
}

@ARTICLE{Walch2015,
       author = {{Walch}, S. and {Girichidis}, P. and {Naab}, T. and {Gatto}, A. and {Glover}, S.~C.~O. and {W{\"u}nsch}, R. and {Klessen}, R.~S. and {Clark}, P.~C. and {Peters}, T. and {Derigs}, D. and {Baczynski}, C.},
        title = "{The SILCC (SImulating the LifeCycle of molecular Clouds) project - I. Chemical evolution of the supernova-driven ISM}",
      journal = {\mnras},
         year = 2015,
        month = nov,
       volume = {454},
       number = {1},
        pages = {238-268},
          doi = {10.1093/mnras/stv1975},
archivePrefix = {arXiv},
       eprint = {1412.2749},
 primaryClass = {astro-ph.GA},
       adsurl = {https://ui.adsabs.harvard.edu/abs/2015MNRAS.454..238W}
}

@ARTICLE{Hu2021,
       author = {{Hu}, Chia-Yu and {Sternberg}, Amiel and {van Dishoeck}, Ewine F.},
        title = "{Metallicity Dependence of the H/H$_{2}$ and C$^{+}$/C/CO Distributions in a Resolved Self-regulating Interstellar Medium}",
      journal = {\apj},
         year = 2021,
        month = oct,
       volume = {920},
       number = {1},
          eid = {44},
        pages = {44},
          doi = {10.3847/1538-4357/ac0dbd},
archivePrefix = {arXiv},
       eprint = {2103.03889},
 primaryClass = {astro-ph.GA},
       adsurl = {https://ui.adsabs.harvard.edu/abs/2021ApJ...920...44H}
}

@ARTICLE{Katz2022,
       author = {{Katz}, Harley},
        title = "{RAMSES-RTZ: non-equilibrium metal chemistry and cooling coupled to on-the-fly radiation hydrodynamics}",
      journal = {\mnras},
         year = 2022,
        month = may,
       volume = {512},
       number = {1},
        pages = {348-365},
          doi = {10.1093/mnras/stac423},
archivePrefix = {arXiv},
       eprint = {2202.04083},
 primaryClass = {astro-ph.GA},
       adsurl = {https://ui.adsabs.harvard.edu/abs/2022MNRAS.512..348K}
}

@ARTICLE{Hu2017Grackle,
       author = {{Hu}, Chia-Yu and {Naab}, Thorsten and {Glover}, Simon C.~O. and {Walch}, Stefanie and {Clark}, Paul C.},
        title = "{Variable interstellar radiation fields in simulated dwarf galaxies: supernovae versus photoelectric heating}",
      journal = {\mnras},
         year = 2017,
        month = oct,
       volume = {471},
       number = {2},
        pages = {2151-2173},
          doi = {10.1093/mnras/stx1773},
archivePrefix = {arXiv},
       eprint = {1701.08779},
 primaryClass = {astro-ph.GA},
       adsurl = {https://ui.adsabs.harvard.edu/abs/2017MNRAS.471.2151H}
}

@ARTICLE{Burkhart2025,
       author = {{Burkhart}, Blakesley and {Dharmawardena}, Thavisha E. and {Bialy}, Shmuel and {Haworth}, Thomas J. and {Cruz Aguirre}, Fernando and {Jo}, Young-Soo and {Andersson}, B.-G. and {Chung}, Haeun and {Edelstein}, Jerry and {Grenier}, Isabelle and {Hamden}, Erika T. and {Han}, Wonyong and {Hoadley}, Keri and {Lee}, Min-Young and {Min}, Kyoung-Wook and {M{\"u}ller}, Thomas and {Pattle}, Kate and {Peek}, J.~E.~G. and {Pleiss}, Geoff and {Schiminovich}, David and {Seon}, Kwang-Il and {Wilson}, Andrew Gordon and {Zucker}, Catherine},
        title = "{A nearby dark molecular cloud in the Local Bubble revealed via H$_{2}$ fluorescence}",
      journal = {Nature Astronomy},
         year = 2025,
        month = jul,
       volume = {9},
        pages = {1064-1072},
          doi = {10.1038/s41550-025-02541-7},
archivePrefix = {arXiv},
       eprint = {2504.17843},
 primaryClass = {astro-ph.GA},
       adsurl = {https://ui.adsabs.harvard.edu/abs/2025NatAs...9.1064B}
}

@ARTICLE{Ploeckinger2025,
       author = {{Ploeckinger}, Sylvia and {Richings}, Alexander J. and {Schaye}, Joop and {Trayford}, James W. and {Schaller}, Matthieu and {Chaikin}, Evgenii},
        title = "{HYBRID-CHIMES: a model for radiative cooling and the abundances of ions and molecules in simulations of galaxy formation}",
      journal = {\mnras},
         year = 2025,
        month = oct,
       volume = {543},
       number = {2},
        pages = {891-916},
          doi = {10.1093/mnras/staf1402},
archivePrefix = {arXiv},
       eprint = {2506.15773},
 primaryClass = {astro-ph.GA},
       adsurl = {https://ui.adsabs.harvard.edu/abs/2025MNRAS.543..891P}
}

@ARTICLE{cloudy2025,
       author = {{Gunasekera}, Chamani M. and {van Hoof}, Peter A.~M. and {Dehghanian}, Maryam and {Chakraborty}, Priyanka and {Shaw}, Gargi and {Bianchi}, Stefano and {Chatzikos}, Marios and {Tsujimoto}, Masahiro and {Ferland}, Gary J.},
        title = "{The 2025 Release of Cloudy}",
      journal = {arXiv e-prints},
         year = 2025,
        month = aug,
          eid = {arXiv:2508.01102},
        pages = {arXiv:2508.01102},
          doi = {10.48550/arXiv.2508.01102},
archivePrefix = {arXiv},
       eprint = {2508.01102},
 primaryClass = {astro-ph.GA},
       adsurl = {https://ui.adsabs.harvard.edu/abs/2025arXiv250801102G}
}

@ARTICLE{Emerick2019,
       author = {{Emerick}, Andrew and {Bryan}, Greg L. and {Mac Low}, Mordecai-Mark},
        title = "{Simulating an isolated dwarf galaxy with multichannel feedback and chemical yields from individual stars}",
      journal = {\mnras},
         year = 2019,
        month = jan,
       volume = {482},
       number = {1},
        pages = {1304-1329},
          doi = {10.1093/mnras/sty2689},
archivePrefix = {arXiv},
       eprint = {1807.07182},
 primaryClass = {astro-ph.GA},
       adsurl = {https://ui.adsabs.harvard.edu/abs/2019MNRAS.482.1304E}
}

@ARTICLE{Smith2021,
       author = {{Smith}, Matthew C.},
        title = "{The sensitivity of stellar feedback to IMF averaging versus IMF sampling in galaxy formation simulations}",
      journal = {\mnras},
         year = 2021,
        month = apr,
       volume = {502},
       number = {4},
        pages = {5417-5437},
          doi = {10.1093/mnras/stab291},
archivePrefix = {arXiv},
       eprint = {2010.10533},
 primaryClass = {astro-ph.GA},
       adsurl = {https://ui.adsabs.harvard.edu/abs/2021MNRAS.502.5417S}
}

@ARTICLE{Emerick2018,
       author = {{Emerick}, Andrew and {Bryan}, Greg L. and {Mac Low}, Mordecai-Mark},
        title = "{Stellar Radiation Is Critical for Regulating Star Formation and Driving Outflows in Low-mass Dwarf Galaxies}",
      journal = {\apjl},
         year = 2018,
        month = oct,
       volume = {865},
       number = {2},
          eid = {L22},
        pages = {L22},
          doi = {10.3847/2041-8213/aae315},
archivePrefix = {arXiv},
       eprint = {1808.00468},
 primaryClass = {astro-ph.GA},
       adsurl = {https://ui.adsabs.harvard.edu/abs/2018ApJ...865L..22E}
}

@ARTICLE{Revaz2018,
       author = {{Revaz}, Yves and {Jablonka}, Pascale},
        title = "{Pushing back the limits: detailed properties of dwarf galaxies in a {\ensuremath{\Lambda}}CDM universe}",
      journal = {\aap},
         year = 2018,
        month = aug,
       volume = {616},
          eid = {A96},
        pages = {A96},
          doi = {10.1051/0004-6361/201832669},
archivePrefix = {arXiv},
       eprint = {1801.06222},
 primaryClass = {astro-ph.GA},
       adsurl = {https://ui.adsabs.harvard.edu/abs/2018A&A...616A..96R}
}

@ARTICLE{Gentry2019,
       author = {{Gentry}, Eric S. and {Krumholz}, Mark R. and {Madau}, Piero and {Lupi}, Alessandro},
        title = "{The momentum budget of clustered supernova feedback in a 3D, magnetized medium}",
      journal = {\mnras},
         year = 2019,
        month = mar,
       volume = {483},
       number = {3},
        pages = {3647-3658},
          doi = {10.1093/mnras/sty3319},
archivePrefix = {arXiv},
       eprint = {1802.06860},
 primaryClass = {astro-ph.GA},
       adsurl = {https://ui.adsabs.harvard.edu/abs/2019MNRAS.483.3647G}
}

@ARTICLE{SimbaEOR2024,
       author = {{Jones}, E. and {Smith}, B. and {Dav{\'e}}, R. and {Narayanan}, D. and {Li}, Q.},
        title = "{SIMBA-EOR: early galaxy formation in the SIMBA simulation including a new sub-grid interstellar medium model}",
      journal = {\mnras},
         year = 2024,
        month = dec,
       volume = {535},
       number = {2},
        pages = {1293-1314},
          doi = {10.1093/mnras/stae2445},
archivePrefix = {arXiv},
       eprint = {2402.06728},
 primaryClass = {astro-ph.GA},
       adsurl = {https://ui.adsabs.harvard.edu/abs/2024MNRAS.535.1293J}
}

@ARTICLE{Dave2016,
       author = {{Dav{\'e}}, Romeel and {Thompson}, Robert and {Hopkins}, Philip F.},
        title = "{MUFASA: galaxy formation simulations with meshless hydrodynamics}",
      journal = {\mnras},
         year = 2016,
        month = nov,
       volume = {462},
       number = {3},
        pages = {3265-3284},
          doi = {10.1093/mnras/stw1862},
archivePrefix = {arXiv},
       eprint = {1604.01418},
 primaryClass = {astro-ph.GA},
       adsurl = {https://ui.adsabs.harvard.edu/abs/2016MNRAS.462.3265D}
}

@ARTICLE{Krumholz2009,
       author = {{Krumholz}, Mark R. and {McKee}, Christopher F. and {Tumlinson}, Jason},
        title = "{The Atomic-to-Molecular Transition in Galaxies. II: H I and H$_{2}$ Column Densities}",
      journal = {\apj},
         year = 2009,
        month = mar,
       volume = {693},
       number = {1},
        pages = {216-235},
          doi = {10.1088/0004-637X/693/1/216},
archivePrefix = {arXiv},
       eprint = {0811.0004},
 primaryClass = {astro-ph},
       adsurl = {https://ui.adsabs.harvard.edu/abs/2009ApJ...693..216K}
}

@ARTICLE{Fryxell2000,
       author = {{Fryxell}, B. and {Olson}, K. and {Ricker}, P. and {Timmes}, F.~X. and {Zingale}, M. and {Lamb}, D.~Q. and {MacNeice}, P. and {Rosner}, R. and {Truran}, J.~W. and {Tufo}, H.},
        title = "{FLASH: An Adaptive Mesh Hydrodynamics Code for Modeling Astrophysical Thermonuclear Flashes}",
      journal = {\apjs},
         year = 2000,
        month = nov,
       volume = {131},
       number = {1},
        pages = {273-334},
          doi = {10.1086/317361},
       adsurl = {https://ui.adsabs.harvard.edu/abs/2000ApJS..131..273F}
}

@ARTICLE{Micic2012,
       author = {{Micic}, Milica and {Glover}, Simon C.~O. and {Federrath}, Christoph and {Klessen}, Ralf S.},
        title = "{Modelling H$_{2}$ formation in the turbulent interstellar medium: solenoidal versus compressive turbulent forcing}",
      journal = {\mnras},
         year = 2012,
        month = apr,
       volume = {421},
       number = {3},
        pages = {2531-2542},
          doi = {10.1111/j.1365-2966.2012.20477.x},
archivePrefix = {arXiv},
       eprint = {1103.3056},
 primaryClass = {astro-ph.SR},
       adsurl = {https://ui.adsabs.harvard.edu/abs/2012MNRAS.421.2531M}
}

@ARTICLE{Hu2016,
       author = {{Hu}, Chia-Yu and {Naab}, Thorsten and {Walch}, Stefanie and {Glover}, Simon C.~O. and {Clark}, Paul C.},
        title = "{Star formation and molecular hydrogen in dwarf galaxies: a non-equilibrium view}",
      journal = {\mnras},
         year = 2016,
        month = jun,
       volume = {458},
       number = {4},
        pages = {3528-3553},
          doi = {10.1093/mnras/stw544},
archivePrefix = {arXiv},
       eprint = {1510.05644},
 primaryClass = {astro-ph.GA},
       adsurl = {https://ui.adsabs.harvard.edu/abs/2016MNRAS.458.3528H}
}

@ARTICLE{Lahen2020,
       author = {{Lah{\'e}n}, Natalia and {Naab}, Thorsten and {Johansson}, Peter H. and {Elmegreen}, Bruce and {Hu}, Chia-Yu and {Walch}, Stefanie and {Steinwandel}, Ulrich P. and {Moster}, Benjamin P.},
        title = "{The GRIFFIN Project{\textemdash}Formation of Star Clusters with Individual Massive Stars in a Simulated Dwarf Galaxy Starburst}",
      journal = {\apj},
         year = 2020,
        month = mar,
       volume = {891},
       number = {1},
          eid = {2},
        pages = {2},
          doi = {10.3847/1538-4357/ab7190},
archivePrefix = {arXiv},
       eprint = {1911.05093},
 primaryClass = {astro-ph.GA},
       adsurl = {https://ui.adsabs.harvard.edu/abs/2020ApJ...891....2L}
}

@ARTICLE{Teyssier2002,
       author = {{Teyssier}, R.},
        title = "{Cosmological hydrodynamics with adaptive mesh refinement. A new high resolution code called RAMSES}",
      journal = {\aap},
         year = 2002,
        month = apr,
       volume = {385},
        pages = {337-364},
          doi = {10.1051/0004-6361:20011817},
archivePrefix = {arXiv},
       eprint = {astro-ph/0111367},
 primaryClass = {astro-ph},
       adsurl = {https://ui.adsabs.harvard.edu/abs/2002A&A...385..337T}
}

@ARTICLE{Rosdahl2015,
       author = {{Rosdahl}, J. and {Teyssier}, R.},
        title = "{A scheme for radiation pressure and photon diffusion with the M1 closure in RAMSES-RT}",
      journal = {\mnras},
         year = 2015,
        month = jun,
       volume = {449},
       number = {4},
        pages = {4380-4403},
          doi = {10.1093/mnras/stv567},
archivePrefix = {arXiv},
       eprint = {1411.6440},
 primaryClass = {astro-ph.IM},
       adsurl = {https://ui.adsabs.harvard.edu/abs/2015MNRAS.449.4380R}
}

@ARTICLE{Schaye2025colibre,
       author = {{Schaye}, Joop and {Chaikin}, Evgenii and {Schaller}, Matthieu and {Ploeckinger}, Sylvia and {Hu{\v{s}}ko}, Filip and {McGibbon}, Rob and {Trayford}, James W. and {Ben{\'\i}tez-Llambay}, Alejandro and {Correa}, Camila and {Frenk}, Carlos S. and {Richings}, Alexander J. and {Forouhar Moreno}, Victor J. and {Bah{\'e}}, Yannick M. and {Borrow}, Josh and {Durrant}, Anna and {Gebek}, Andrea and {Helly}, John C. and {Jenkins}, Adrian and {Lacey}, Cedric G. and {Ludlow}, Aaron and {Nobels}, Folkert S.~J.},
        title = "{The COLIBRE project: cosmological hydrodynamical simulations of galaxy formation and evolution}",
      journal = {arXiv e-prints},
         year = 2025,
        month = aug,
          eid = {arXiv:2508.21126},
        pages = {arXiv:2508.21126},
          doi = {10.48550/arXiv.2508.21126},
archivePrefix = {arXiv},
       eprint = {2508.21126},
 primaryClass = {astro-ph.GA},
       adsurl = {https://ui.adsabs.harvard.edu/abs/2025arXiv250821126S}
}

@ARTICLE{Chaikin2025a,
       author = {{Chaikin}, Evgenii and {Schaye}, Joop and {Schaller}, Matthieu and {Ploeckinger}, Sylvia and {Bah{\'e}}, Yannick M. and {Ben{\'\i}tez-Llambay}, Alejandro and {Correa}, Camila and {Forouhar Moreno}, Victor J. and {Frenk}, Carlos S. and {Hu{\v{s}}ko}, Filip and {Kugel}, Roi and {McGibbon}, Robert and {Richings}, Alexander J. and {Trayford}, James W. and {Borrow}, Josh and {Crain}, Robert A. and {Helly}, John C. and {Lacey}, Cedric G. and {Ludlow}, Aaron and {Nobels}, Folkert S.~J.},
        title = "{COLIBRE: calibrating subgrid feedback in cosmological simulations that include a cold gas phase}",
      journal = {arXiv e-prints},
         year = 2025,
        month = sep,
          eid = {arXiv:2509.04067},
        pages = {arXiv:2509.04067},
          doi = {10.48550/arXiv.2509.04067},
archivePrefix = {arXiv},
       eprint = {2509.04067},
 primaryClass = {astro-ph.GA},
       adsurl = {https://ui.adsabs.harvard.edu/abs/2025arXiv250904067C}
}

@ARTICLE{Katz2017,
       author = {{Katz}, Harley and {Kimm}, Taysun and {Sijacki}, Debora and {Haehnelt}, Martin G.},
        title = "{Interpreting ALMA observations of the ISM during the epoch of reionization}",
      journal = {\mnras},
         year = 2017,
        month = jul,
       volume = {468},
       number = {4},
        pages = {4831-4861},
          doi = {10.1093/mnras/stx608},
archivePrefix = {arXiv},
       eprint = {1612.01786},
 primaryClass = {astro-ph.GA},
       adsurl = {https://ui.adsabs.harvard.edu/abs/2017MNRAS.468.4831K}
}

@ARTICLE{Zier2025,
       author = {{Zier}, Oliver and {Kannan}, Rahul and {Smith}, Aaron and {Puchwein}, Ewald and {Vogelsberger}, Mark and {Borrow}, Josh and {Garaldi}, Enrico and {Keating}, Laura and {McClymont}, William and {Shen}, Xuejian and {Hernquist}, Lars},
        title = "{The THESAN-ZOOM project: Population III star formation continues until the end of reionization}",
      journal = {\mnras},
         year = 2025,
        month = nov,
       volume = {544},
       number = {1},
        pages = {410-429},
          doi = {10.1093/mnras/staf1053},
archivePrefix = {arXiv},
       eprint = {2503.03806},
 primaryClass = {astro-ph.GA},
       adsurl = {https://ui.adsabs.harvard.edu/abs/2025MNRAS.544..410Z}
}

@ARTICLE{Kannan2025,
       author = {{Kannan}, Rahul and {Puchwein}, Ewald and {Smith}, Aaron and {Borrow}, Josh and {Garaldi}, Enrico and {Keating}, Laura and {Vogelsberger}, Mark and {Zier}, Oliver and {McClymont}, William and {Shen}, Xuejian and {Popovic}, Filip and {Tacchella}, Sandro and {Hernquist}, Lars and {Springel}, Volker},
        title = "{Introducing the THESAN-ZOOM project: radiation-hydrodynamic simulations of high-redshift galaxies with a multi-phase interstellar medium}",
      journal = {The Open Journal of Astrophysics},
         year = 2025,
        month = oct,
       volume = {8},
          eid = {153},
        pages = {153},
          doi = {10.33232/001c.145804},
archivePrefix = {arXiv},
       eprint = {2502.20437},
 primaryClass = {astro-ph.GA},
       adsurl = {https://ui.adsabs.harvard.edu/abs/2025OJAp....8E.153K}
}

@ARTICLE{Kannan2020,
       author = {{Kannan}, Rahul and {Marinacci}, Federico and {Vogelsberger}, Mark and {Sales}, Laura V. and {Torrey}, Paul and {Springel}, Volker and {Hernquist}, Lars},
        title = "{Simulating the interstellar medium of galaxies with radiative transfer, non-equilibrium thermochemistry, and dust}",
      journal = {\mnras},
         year = 2020,
        month = dec,
       volume = {499},
       number = {4},
        pages = {5732-5748},
          doi = {10.1093/mnras/staa3249},
archivePrefix = {arXiv},
       eprint = {1910.14041},
 primaryClass = {astro-ph.GA},
       adsurl = {https://ui.adsabs.harvard.edu/abs/2020MNRAS.499.5732K}
}

@ARTICLE{Shaw2005,
       author = {{Shaw}, G. and {Ferland}, G.~J. and {Abel}, N.~P. and {Stancil}, P.~C. and {van Hoof}, P.~A.~M.},
        title = "{Molecular Hydrogen in Star-forming Regions: Implementation of its Microphysics in CLOUDY}",
      journal = {\apj},
         year = 2005,
        month = may,
       volume = {624},
       number = {2},
        pages = {794-807},
          doi = {10.1086/429215},
archivePrefix = {arXiv},
       eprint = {astro-ph/0501485},
 primaryClass = {astro-ph},
       adsurl = {https://ui.adsabs.harvard.edu/abs/2005ApJ...624..794S}
}

@ARTICLE{Changa2015,
       author = {{Menon}, Harshitha and {Wesolowski}, Lukasz and {Zheng}, Gengbin and {Jetley}, Pritish and {Kale}, Laxmikant and {Quinn}, Thomas and {Governato}, Fabio},
        title = "{Adaptive techniques for clustered N-body cosmological simulations}",
      journal = {Computational Astrophysics and Cosmology},
         year = 2015,
        month = mar,
       volume = {2},
          eid = {1},
        pages = {1},
          doi = {10.1186/s40668-015-0007-9},
archivePrefix = {arXiv},
       eprint = {1409.1929},
 primaryClass = {astro-ph.IM},
       adsurl = {https://ui.adsabs.harvard.edu/abs/2015ComAC...2....1M}
}

@ARTICLE{Cholla2015,
       author = {{Schneider}, Evan E. and {Robertson}, Brant E.},
        title = "{CHOLLA: A New Massively Parallel Hydrodynamics Code for Astrophysical Simulation}",
      journal = {\apjs},
         year = 2015,
        month = apr,
       volume = {217},
       number = {2},
          eid = {24},
        pages = {24},
          doi = {10.1088/0067-0049/217/2/24},
archivePrefix = {arXiv},
       eprint = {1410.4194},
 primaryClass = {astro-ph.IM},
       adsurl = {https://ui.adsabs.harvard.edu/abs/2015ApJS..217...24S}
}

@ARTICLE{Enzo2014,
       author = {{Bryan}, Greg L. and {Norman}, Michael L. and {O'Shea}, Brian W. and {Abel}, Tom and {Wise}, John H. and {Turk}, Matthew J. and {Reynolds}, Daniel R. and {Collins}, David C. and {Wang}, Peng and {Skillman}, Samuel W. and {Smith}, Britton and {Harkness}, Robert P. and {Bordner}, James and {Kim}, Ji-hoon and {Kuhlen}, Michael and {Xu}, Hao and {Goldbaum}, Nathan and {Hummels}, Cameron and {Kritsuk}, Alexei G. and {Tasker}, Elizabeth and {Skory}, Stephen and {Simpson}, Christine M. and {Hahn}, Oliver and {Oishi}, Jeffrey S. and {So}, Geoffrey C. and {Zhao}, Fen and {Cen}, Renyue and {Li}, Yuan and {Enzo Collaboration}},
        title = "{ENZO: An Adaptive Mesh Refinement Code for Astrophysics}",
      journal = {\apjs},
         year = 2014,
        month = apr,
       volume = {211},
       number = {2},
          eid = {19},
        pages = {19},
          doi = {10.1088/0067-0049/211/2/19},
archivePrefix = {arXiv},
       eprint = {1307.2265},
 primaryClass = {astro-ph.IM},
       adsurl = {https://ui.adsabs.harvard.edu/abs/2014ApJS..211...19B}
}

@ARTICLE{Gasoline2017,
       author = {{Wadsley}, James W. and {Keller}, Benjamin W. and {Quinn}, Thomas R.},
        title = "{Gasoline2: a modern smoothed particle hydrodynamics code}",
      journal = {\mnras},
         year = 2017,
        month = oct,
       volume = {471},
       number = {2},
        pages = {2357-2369},
          doi = {10.1093/mnras/stx1643},
archivePrefix = {arXiv},
       eprint = {1707.03824},
 primaryClass = {astro-ph.IM},
       adsurl = {https://ui.adsabs.harvard.edu/abs/2017MNRAS.471.2357W}
}

@ARTICLE{Gizmo2015,
       author = {{Hopkins}, Philip F.},
        title = "{A new class of accurate, mesh-free hydrodynamic simulation methods}",
      journal = {\mnras},
         year = 2015,
        month = jun,
       volume = {450},
       number = {1},
        pages = {53-110},
          doi = {10.1093/mnras/stv195},
archivePrefix = {arXiv},
       eprint = {1409.7395},
 primaryClass = {astro-ph.CO},
       adsurl = {https://ui.adsabs.harvard.edu/abs/2015MNRAS.450...53H}
}

@ARTICLE{Arepo2020,
       author = {{Weinberger}, Rainer and {Springel}, Volker and {Pakmor}, R{\"u}diger},
        title = "{The AREPO Public Code Release}",
      journal = {\apjs},
         year = 2020,
        month = jun,
       volume = {248},
       number = {2},
          eid = {32},
        pages = {32},
          doi = {10.3847/1538-4365/ab908c},
archivePrefix = {arXiv},
       eprint = {1909.04667},
 primaryClass = {astro-ph.IM},
       adsurl = {https://ui.adsabs.harvard.edu/abs/2020ApJS..248...32W}
}

@ARTICLE{Wolfire2008,
       author = {{Wolfire}, Mark G. and {Tielens}, A.~G.~G.~M. and {Hollenbach}, David and {Kaufman}, M.~J.},
        title = "{Chemical Rates on Small Grains and PAHs: C$^{+}$ Recombination and H$_{2}$ Formation}",
      journal = {\apj},
         year = 2008,
        month = jun,
       volume = {680},
       number = {1},
        pages = {384-397},
          doi = {10.1086/587688},
archivePrefix = {arXiv},
       eprint = {0803.0138},
 primaryClass = {astro-ph},
       adsurl = {https://ui.adsabs.harvard.edu/abs/2008ApJ...680..384W}
}

@ARTICLE{Bolatto2011,
       author = {{Bolatto}, Alberto D. and {Leroy}, Adam K. and {Jameson}, Katherine and {Ostriker}, Eve and {Gordon}, Karl and {Lawton}, Brandon and {Stanimirovi{\'c}}, Sne{\v{z}}ana and {Israel}, Frank P. and {Madden}, Suzanne C. and {Hony}, Sacha and {Sandstrom}, Karin M. and {Bot}, Caroline and {Rubio}, M{\'o}nica and {Winkler}, P. Frank and {Roman-Duval}, Julia and {van Loon}, Jacco Th. and {Oliveira}, Joana M. and {Indebetouw}, R{\'e}my},
        title = "{The State of the Gas and the Relation between Gas and Star Formation at Low Metallicity: The Small Magellanic Cloud}",
      journal = {\apj},
         year = 2011,
        month = nov,
       volume = {741},
       number = {1},
          eid = {12},
        pages = {12},
          doi = {10.1088/0004-637X/741/1/12},
archivePrefix = {arXiv},
       eprint = {1107.1717},
 primaryClass = {astro-ph.CO},
       adsurl = {https://ui.adsabs.harvard.edu/abs/2011ApJ...741...12B}
}






\bsp	
\label{lastpage}
\end{document}